\documentclass[times]{article}
\author{Christophe Lohou}

\usepackage{arxiv}
\usepackage[utf8]{inputenc} 
\usepackage[T1]{fontenc}    
\usepackage[hidelinks]{hyperref}       
\usepackage{url}            
\usepackage{booktabs}       
\usepackage{amsfonts}       
\usepackage{nicefrac}       
\usepackage{microtype}      
\usepackage{cleveref}       
\usepackage{graphicx}
\usepackage{doi}
\usepackage{longtable}
\usepackage{supertabular}
\usepackage[english]{babel}
\usepackage[table,xcdraw]{xcolor}
\usepackage{breakurl}

\usepackage{amssymb}
\usepackage{latexsym}

\usepackage{orcidlink} 
\usepackage{multirow}
\usepackage{lscape}

\graphicspath{ {./} }
\begin{document}

\title{Segmentation of separated Lumens\\ in 3D CTA images of Aortic Dissection}
\author{
\textbf{Christophe Lohou} \orcidlink{0000-0001-5352-8237}, \\
    Université Clermont Auvergne,\\
    Clermont Auvergne INP,\\
    CNRS, Institut Pascal\\
    F-63000 Clermont-Ferrand, France\\
	\texttt{christophe.lohou@uca.fr}
\And
\textbf{Bruno Miguel},\\
Université Clermont Auvergne,\\
    Clermont Auvergne INP,\\
    CHU Clermont-Ferrand,\\
    CNRS, Institut Pascal\\
    F-63000 Clermont-Ferrand, France
}

\renewcommand{\headeright}{}
\renewcommand{\undertitle}{}
\renewcommand{\shorttitle}{}

\date{\today}
\maketitle
\begin{abstract}
Aortic dissection is a serious pathology and requires an emergency management. 
It is characterized by one or more \textit{tears}  of the intimal wall of the normal blood duct of the aorta (\textit{true lumen}); 
the blood under pressure then creates a second blood lumen (\textit{false lumen}) in the media tissue. 
The two lumens are separated by an intimal wall, called \textit{flap}. 
From the segmentation of connected lumens (more precisely, blood inside lumens) of an aortic dissection 3D Computed Tomography Angiography (CTA) image, 
our previous studies allow us to retrieve the intimal flap by using Mathematical Morphology operators, 
and characterize intimal tears by 3d thin surfaces that fill them, 
these surfaces are obtained by operating the Aktouf et al. closing algorithm proposed in the framework of Digital Topology. 
Indeed, intimal tears are 3D holes in the intimal flap; 
although it is impossible to directly segment such non-concrete data, 
it is nevertheless possible to “materialize” them with these 3D filling surfaces that may be quantified or make easier the visualization of these holes. 

In this paper, we use these surfaces that fill tears to cut connections between lumens in order to separate them. 

This is the first time that surfaces filling tears are used as an image processing operator 
(to disconnect several parts of a 3D object). 
This lumen separation allows us to provide one of the first cartographies of an aortic dissection, 
that may better visually assist physicians during their diagnosis.

Our method is able to disconnect lumens, that may also lead to enhance several current investigations (registration, segmentation, hemodynamics).

\end{abstract}

\providecommand{\keywords}[1]{%
  \vspace{1em}
  \begin{center}
    \textbf{Mots-clés :}\\[0.5em]
    \begin{tabular}{c}
      #1
    \end{tabular}
  \end{center}
}

\keywords{ segmentation, aortic dissection cartography, 
lumen separation, holes closing algorithm, Digital Topology }

\vspace{1em}
\begin{center}
    \fbox{%
        \begin{minipage}{\linewidth}
        \centering
        \includegraphics[width=0.95\linewidth]{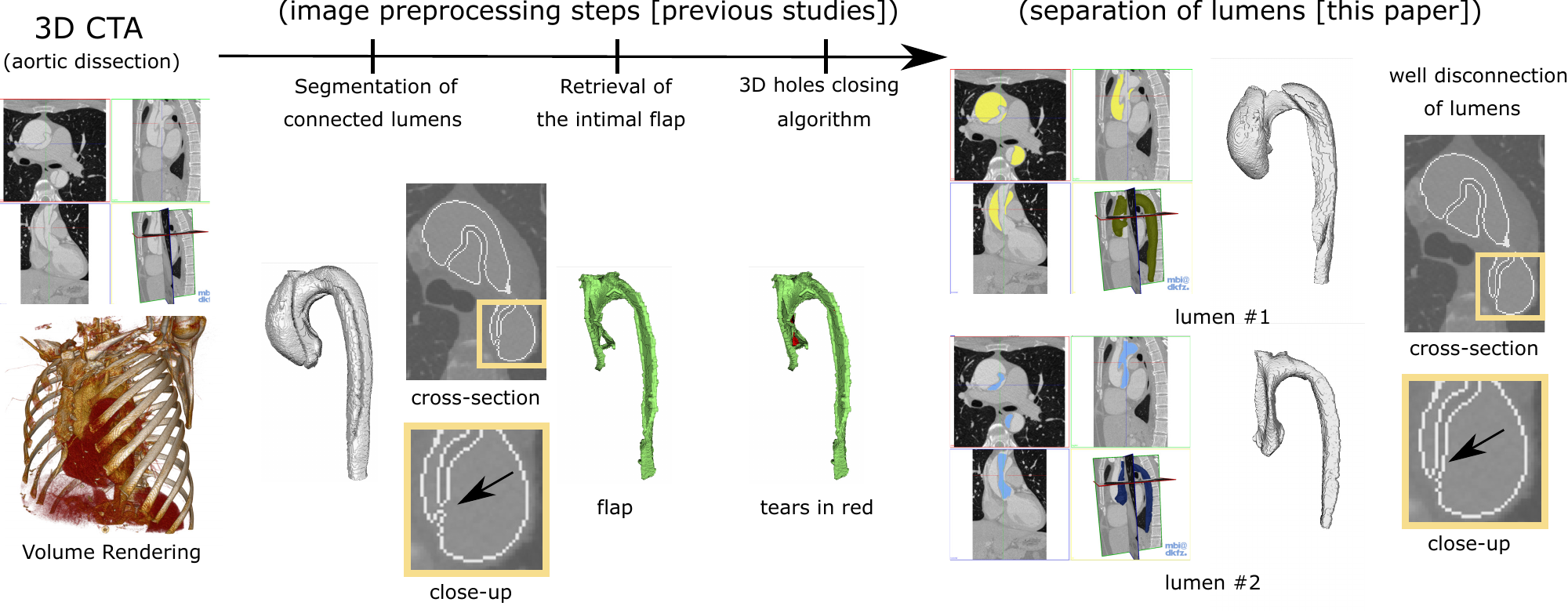}     
        \vspace{0.5em}\\
        \textbf{Graphical Abstract.} Visual overview of the proposed post-processing method to separate true and false lumens from an initial joint segmentation in aortic dissection CT scans.
        \end{minipage}
    }
\end{center}
\vspace{1.5em}

\section{Introduction}
 
\begin{figure}[t!]
\begin{center}
\begin{tabular}{cc}
\includegraphics[height=4.2cm]{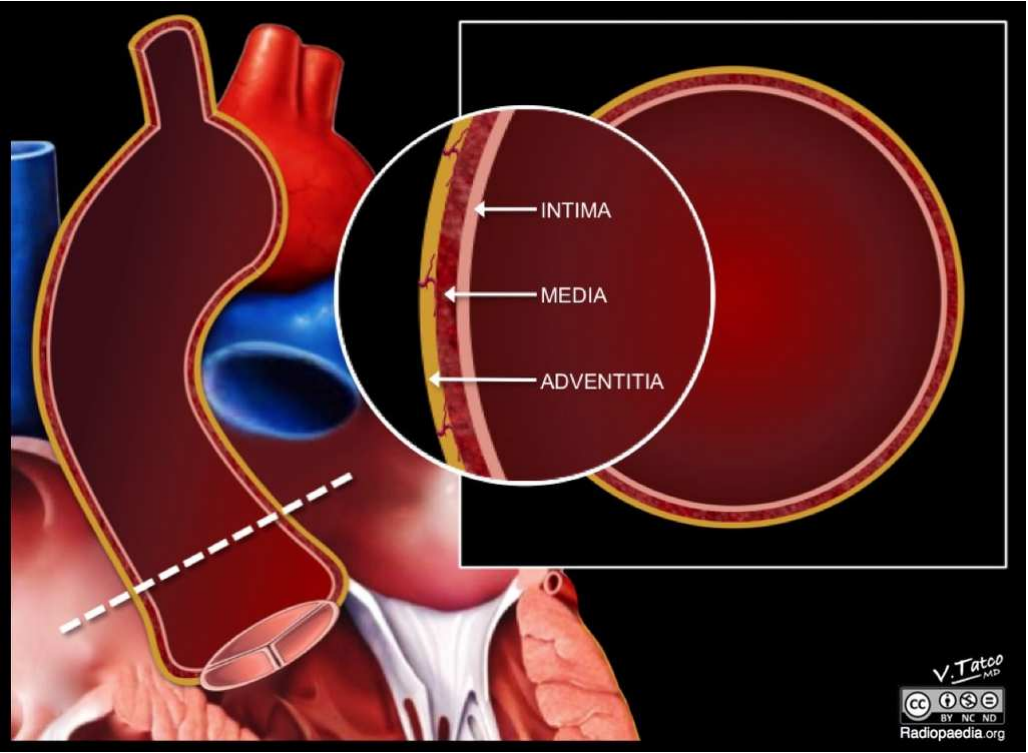} & \includegraphics[height=4.2cm]{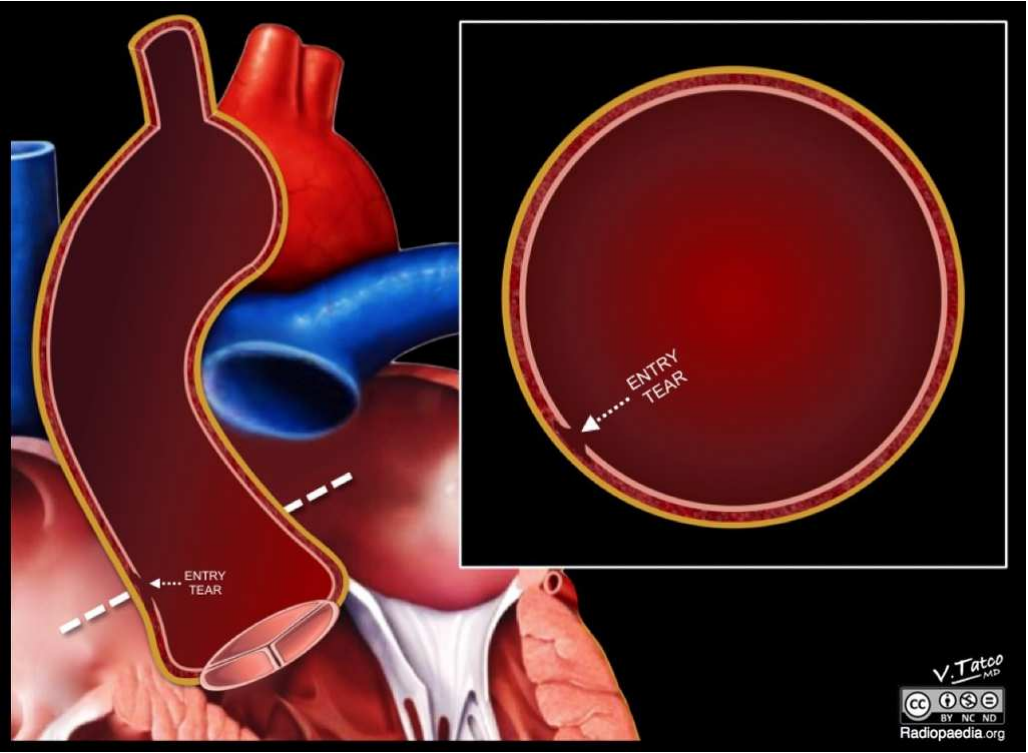}\\
(a) & (b)\\
\multicolumn{2}{c}{\includegraphics[height=4.2cm]{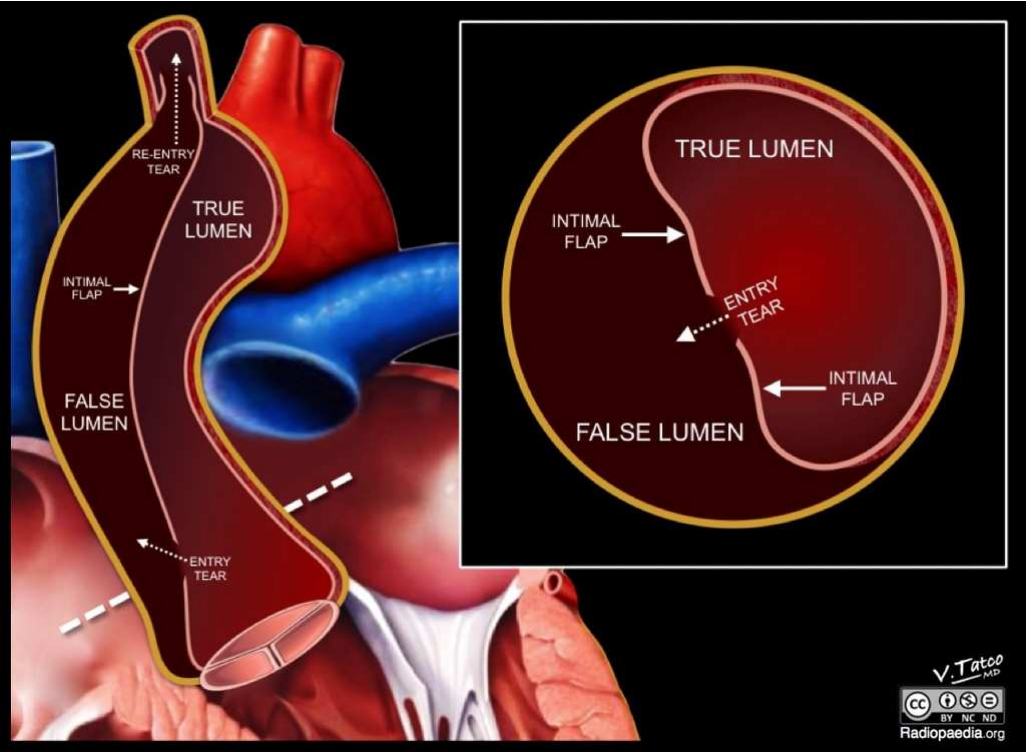}} \\
\multicolumn{2}{c}{(c)}\\
\end{tabular}
\end{center}
\caption{Aortic dissection features: (a) tissue layers, (b) entry tear in the intimal layer, (c) false lumen and intimal flap \cite{Tatco}.}
\label{fig:fig1}
\end{figure}

Aortic dissection is a serious pathology and requires emergency management. It is characterized by one or more \textit{tears} (Fig. \ref{fig:fig1} (b)) of the intimal wall of the normal blood duct of the aorta (\textit{true lumen}); the blood under pressure then creates a second lumen (\textit{false lumen}) (Fig. \ref{fig:fig1} (c)) in the media tissue (Fig. \ref{fig:fig1} (a)). Medical imaging management begins with the acquisition of a patient's 3D Computed Tomography Angiography (3D-CTA) image to visualize the extent of the dissection and the location of the main intimal tears; these characteristics as well as the medical state of the patient allow physicians to decide on the treatment to be put in place (medication, open surgery or interventional radiology with stent-graft placement). 

In previous studies, we first have described several image processing steps to retrieve the external contour of the aorta, then the connected lumens (in fact, blood inside lumens) and the \textit{flap} (intimal wall which is between the two connected lumens, see Fig. \ref{fig:fig1} (c)) \cite{Lohou2013a}. Then, because intimal tears may be considered as holes in the intimal flap, we have performed the Aktouf et al. holes closing algorithm \cite{Aktouf2002} - proposed in the context of Digital Topology - on the intimal flap to add one thin 3D surface by each 3D intimal tear \cite{Lohou2011}. We highlight that conventional image processing opearators cannot segment such an immaterial data (in contrary to concrete organs). 

Few studies have been proposed on the segmentation of aortic dissections: some of them have focused on the extraction of the flap \cite{Lohou2013a, Kovacs2006, Krissian2014}; concerning the intimal tears - as written before, tears correspond to non-concrete data and therefore it is difficult to characterize them -, two studies lead about their visualization by virtual angioscopy \cite{Maldjian2012}, or by using finite elements \cite{Hossien2015}. 
At our knowledge, except for a manual and tedious 3D delineation for all 2D slices, only our previous work \cite{Lohou2011} leads to a semi-automatic characterization of tears by adding surfaces on them. 
Segmentation of lumens includes investigations, either for 2D images, for example by using a multi-scale Wavelet analysis coupled with several models matching (SVM, probabilistic model) \cite{Lee2008}, or for 3D images, for example, by delineating lumens all $3$ slices in a study about the lumens volume variation after an endovascular treatment with a stent-graft \cite{Mellissano2012}, 
or by first defining a $3$D centerline by lumen then by operating an automatic $2$D level-sets segmentation in several $2$D cross-sections orthogonal at centerlines in order to study aorta movements during cardiac and respiratory movements \cite{Suh2014}. 
More recently, a deep convolutional neural network has been proposed with a training database made of $254$ images to segment separated lumens \cite{Li2018}. 
Note also several hemodynamic simulations that do/must not separate the lumens but that propose an interesting visualization of aortic dissection features: 
for example, in Ref. \cite{Chen2013}, a specific model provides detailed flow information of blood transport within the true and false lumens; 
in Ref. \cite{Menichini2016}, a hemodynamics-based model is applied to predict false lumen thrombosis at eight follow-up scans over $3$ years; 
in Ref. \cite{DillonMurphy2016}, a visualization of pressure, wall shear stress and velocity is proposed for connected lumens. 

In this paper, we describe how to simply compute the image of separated lumens, 
once images of connected lumens, flap, tears are precomputed, 
by removing surfaces filling tears from the image of connected lumens. 
In such a way, we may provide one of the first cartographies of an aortic dissection. This is also the first time that surfaces that close $3$D holes are used as an image processing operator to disconnect several parts of a 3D object.
In Sect. \ref{sec:context}, we argue that the simplest way to obtain the two separated lumens is to first get the two lumens connected then to separate them (rather than trying to get them separately). We also quickly recall the algorithm for obtaining surfaces filling holes in 3D objects. In Sect. \ref{sec:our_contribution}, we present our method that consists of removing the image of the intimal tears from the one of the connected lumens to simply disconnect them. In Sect. \ref{sec:method}, we recall the sequence of image processing steps required before this lumens separation step. In Sect. \ref{sec:results}, we give the 3D results and precise 2D results on slices of the lumens disconnection for one patient, as well as the cartography of its dissection. We also give results of the disconnection of lumens for two other images. Finally, we conclude on the possible exploitation of these separated lumens.
 
\section{Context}\label{sec:context}

\subsection{About disconnecting lumens}

We review various possibilities considered for lumen separation. We first consider a direct separation, \textit{i.e.}, the lumens are independently segmented 
(Sect. \ref{sec:direct_separation}), then we study an indirect separation, \textit{i.e.}, 
the lumens are first segmented together and then we attempt to separate them (Sect \ref{sec:indirect_separation}).

\subsubsection{Direct separation}\label{sec:direct_separation}

We present two possible ways in order to directly obtain, by a segmentation process, lumen by lumen, either by an iconic approach (based on intensities) or by a geometric approach.

\begin{itemize}
\item {Iconic approch.} 
For example, we can use a region growing method \cite{Adams94} based on gray level. Such a method usually consists in interactively defining a seed in a lumen, assigning a label to it, and then propagating this label to the neighbors of the seed if they have an intensity in a certain interval with respect to the intensity of the clicked seed, and so on without considering points already seen. This gives a connected component including the clicked seed. We can thus click on several seeds and compute the union of the connected components obtained so as to try to extract one whole lumen; we start again in the second lumen with a new label; then, we may eliminate the common parts of these two segmentations by performing an "or exclusive - XOR" on the labels to get the two independent lumens. This approach seems to be not satisfactory in the areas of intimal tears due to the mixture of blood gray level intensity between lumens.

\begin{figure}[t!]
\begin{center}
\begin{tabular}{cc}
\includegraphics[height=4.7cm]{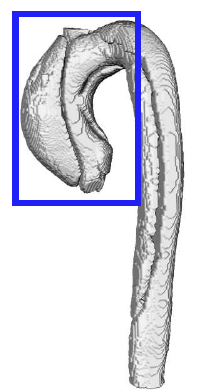} & \includegraphics[height=4.7cm]{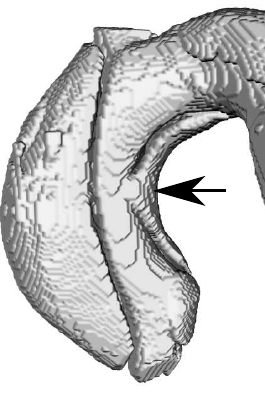} \\
(a) & (b)\\
\end{tabular}
\end{center}
\caption{Intimal tear on the edge of the lumens: (a) initial object, (b) zoom (corresponding to the blue square in (a)). The arrow shows the intimal tear. A segmentation based on curvatures cannot directly obtain separated lumens.}
\label{fig:fig2}
\end{figure}

\item {Geometrical approach.} 
For example, we can perform a segmentation using  deformable models \cite{McInerney96} with the lumen surface curvature as a constraint. 
This implies strong geometrical assumptions about the discontinuity of the curvature of a lumen and that of the surface corresponding to the intimal tear in order to be able to segment lumen by lumen independently. The image of Fig. \ref{fig:fig2} shows a case such that the entry door is on the edge of a lumen and for which there is a too little variation in curvatures; this prevents us from following this possibility to segment lumen by lumen.

\end{itemize}

\subsubsection{Indirect separation}\label{sec:indirect_separation}

In this section, we are dealing with the disconnection of the lumens once they have been segmented together. We propose the analysis of the following 5 cases.

\begin{itemize}
\item Sequence of erosions then dilations of connected lumens.

\begin{figure}[t!]
\begin{center}
\begin{tabular}{ccccc}
\includegraphics[width=2.8cm]{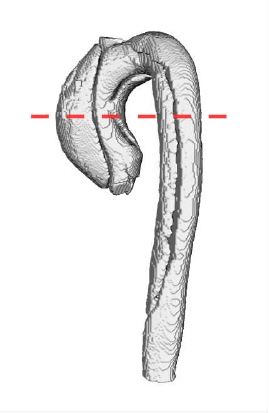} & 
\includegraphics[width=2.8cm]{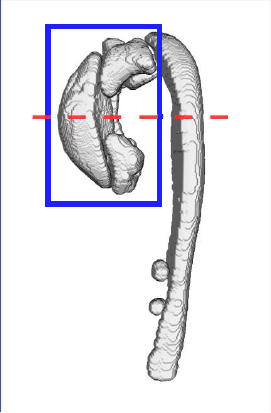} &
\includegraphics[width=2.8cm]{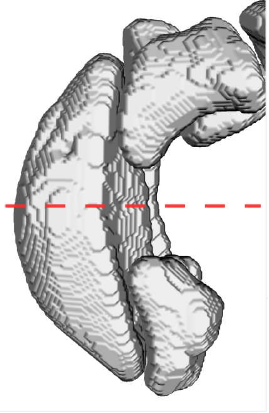} &
\includegraphics[width=2.8cm]{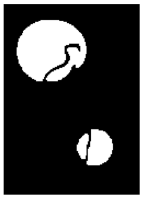} &
\includegraphics[width=2.8cm]{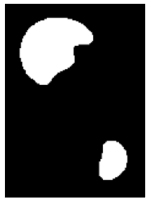}\\
(a) & (b) & (c) & (d) & (e)\\
\end{tabular}
\end{center}
\caption{(a) Initial image, (b) succession of erosions then dilations, (c) zoom (blue rectangle in (b)): a part of one lumen has disappeared. 2D horizontal slices (at the level shown by the orange dotted line in (a-c)): (d) initial slice, (e) slice after the sequence of erosions then dilations:  one lumen has disappeared in this section.}
\label{fig:fig3}
\end{figure}

Intuitively, we can think of first performing a succession of erosions to disconnect the lumens then a succession of dilations - framework of Mathematical Morphology \cite{Serra88} - to try to reconstruct a part of what has been removed. This proposal can not be followed: cf. Fig. \ref{fig:fig3}, the size/height of the connection/tear between the two lumens is greater than the thickness/width of one lumen, therefore this lumen is completely eroded (in several successive slices) before the tear, and it cannot be reconstructed during dilations.  

\item Use of curvature.

The way consisting of following the curvatures of connected lumens would not allow us to disconnect lumens, as previously written (cf. Fig. \ref{fig:fig2}).

\item Use of the flap.

Once the flap has been obtained, we could dilate it so that it could cut both lumens. As for Fig. \ref{fig:fig3}, if one of the two lumens is thinner than the "height" of the tear, dilating the flap a sufficient number of times to divide connections betweens lumens could result in a partial loss of thin lumens. In addition, specific structuring elements should be defined to take into account the non regular curvature of the flap for such dilation. 

\begin{figure}[t!]
\centering
\begin{tabular}{ccc@{\hskip 1cm}c}
\includegraphics[width=3cm]{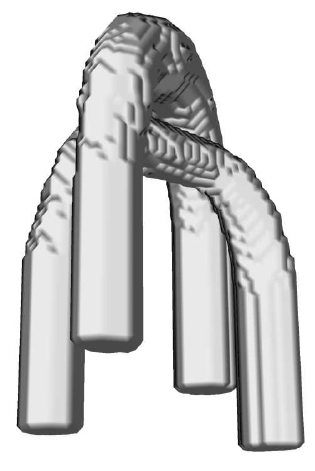} &
\includegraphics[width=3cm]{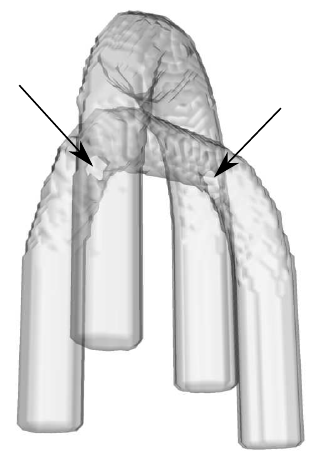} &
\includegraphics[width=3cm]{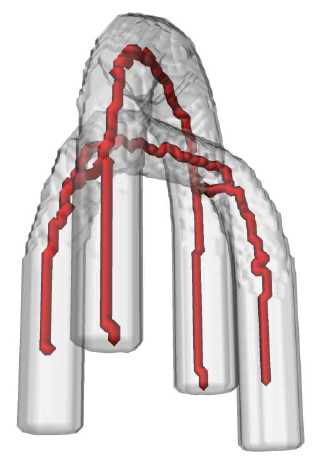} &
\multirow{4}{*}{%
  \begin{minipage}[c]{4cm}
    \vspace*{-2.5cm}  
    \begin{center}
      \includegraphics[width=\linewidth]{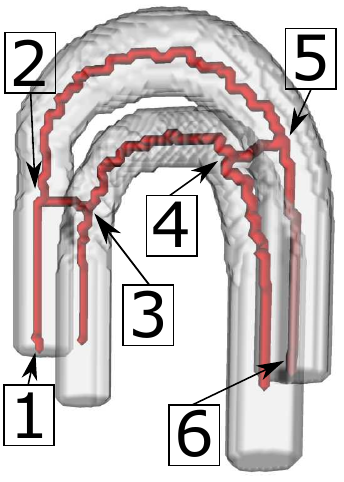}\\
      (g)
    \end{center}
  \end{minipage}
} \\

(a) & (b) & (c) & \\[0.4em]
\includegraphics[width=3cm]{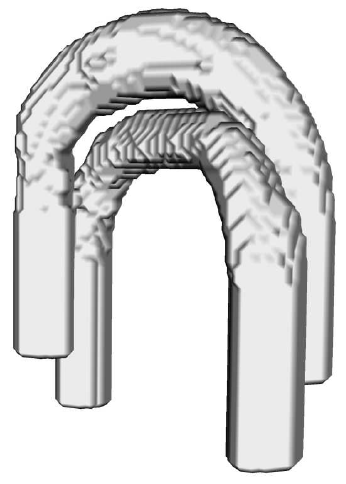} &
\includegraphics[width=3cm]{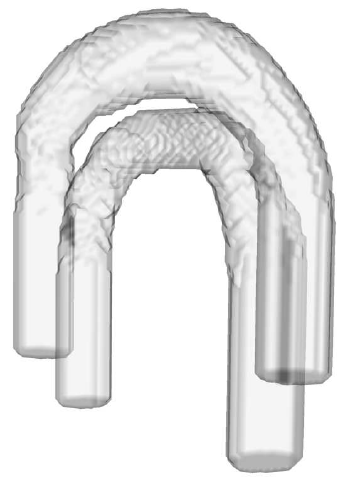} &
\includegraphics[width=3cm]{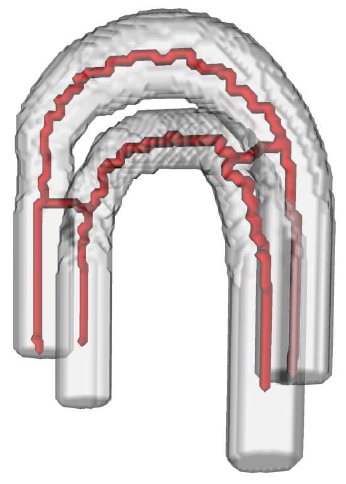} & \\
(d) & (e) & (f) & \\
\end{tabular}

\caption{(a) Initial object, (b) in transparency, arrows represent the location of the two tears, (c) with the skeleton in red (algorithm proposed in Ref. \cite{Lohou2005}); (d-f) with a different viewing orientation. 
(g) The algorithm of the shortest path between points 1 and 6 of a first lumen passes through the second lumen (case of a dissection with two tears).}
\label{fig:fig4}
\end{figure}

\item Use of skeleton.

We might consider the processing of skeletonization/thinning \cite{Kong1989} of connected lumens in order to compute one associated skeleton, 
then by analysis of this skeleton to keep only the centerlines of the skeleton corresponding to each lumen 
(for example, by fixing start and finish points in each lumen and to compute the shortest path  \cite{Dijkstra1959} between these points) 
and finally to operate a conditional dilation \cite{Serra88} of these centerlines inside connected lumens to reconstruct each lumen. 
In fact, each step leads to difficulties. The main problem is that the computing of the shortest path between two points of the same lumen may produce a path that is not only included inside a single lumen but also takes the second lumen through the tears.

This is particularly the case in the object of Fig. \ref{fig:fig4} (obtained from a real image) for which the two lumens are intertwined around two intimal tears : two anchor points (start/arrival) are defined in the first lumen (points 1 and 6); the found path when running the shortest path algorithm starts from point 1 and reaches point 2 in the same first lumen, then crosses at a first tear to reach the second lumen at point 3, then goes to point 4 still in the second lumen, then again crosses a second tear again to reach the first lumen at point 5 and then to reach point 6 of arrival. Here we have the example of a shortest path that changes twice the lumen. 

\begin{figure}[t!]
\begin{center}
\begin{tabular}{ccccc}
\includegraphics[width=2.8cm]{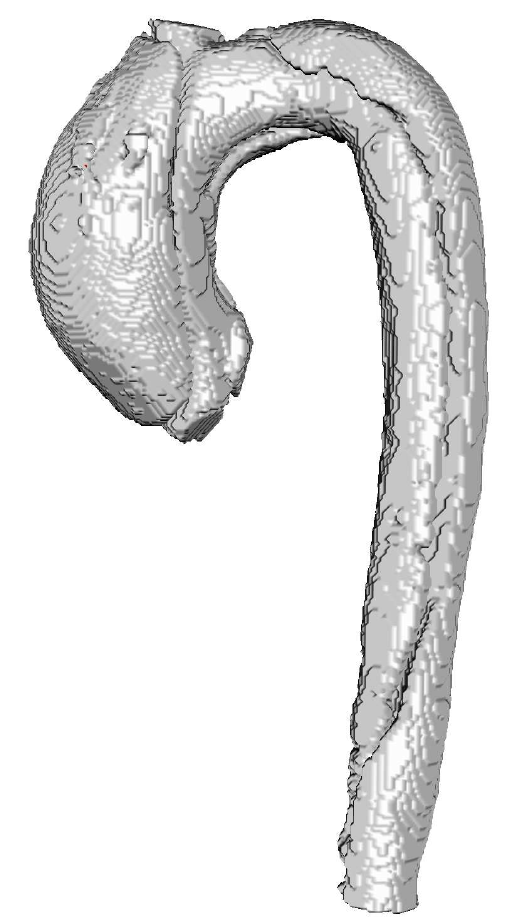} &
\includegraphics[width=2.8cm]{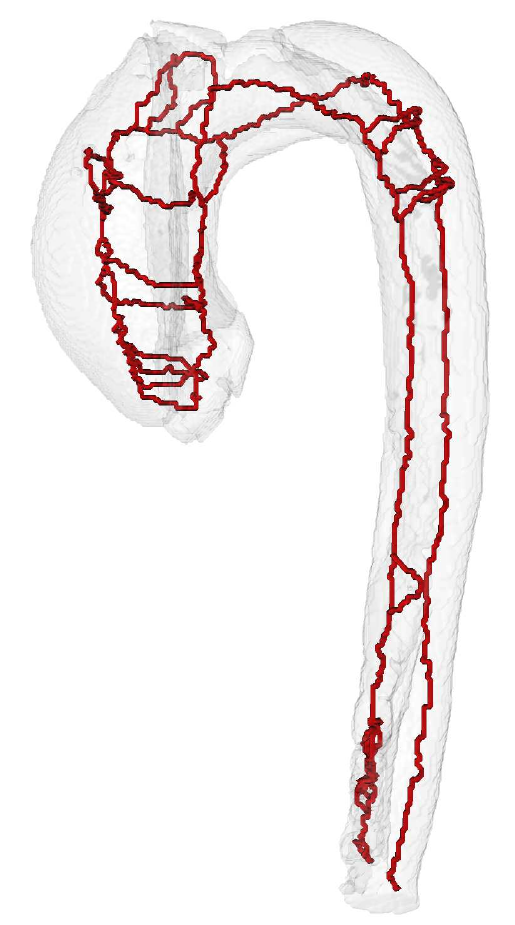} &
\includegraphics[width=2.8cm]{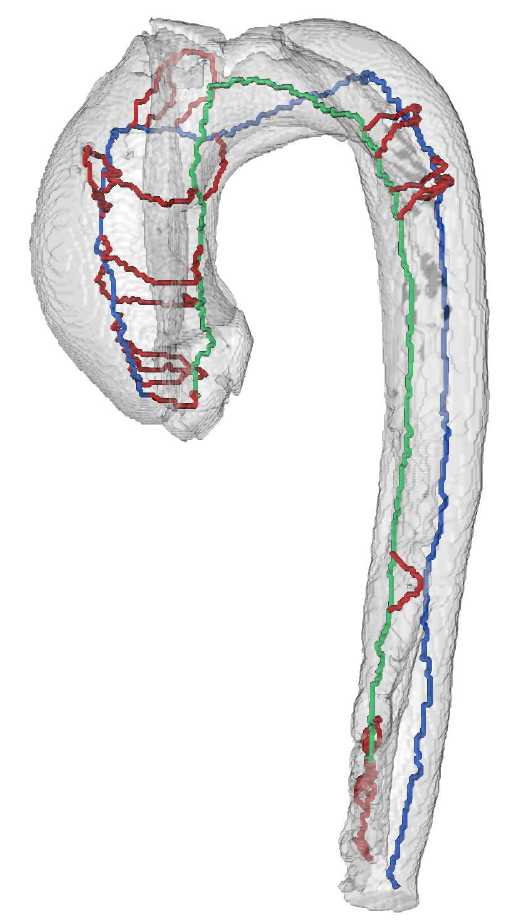} &
\includegraphics[width=2.8cm]{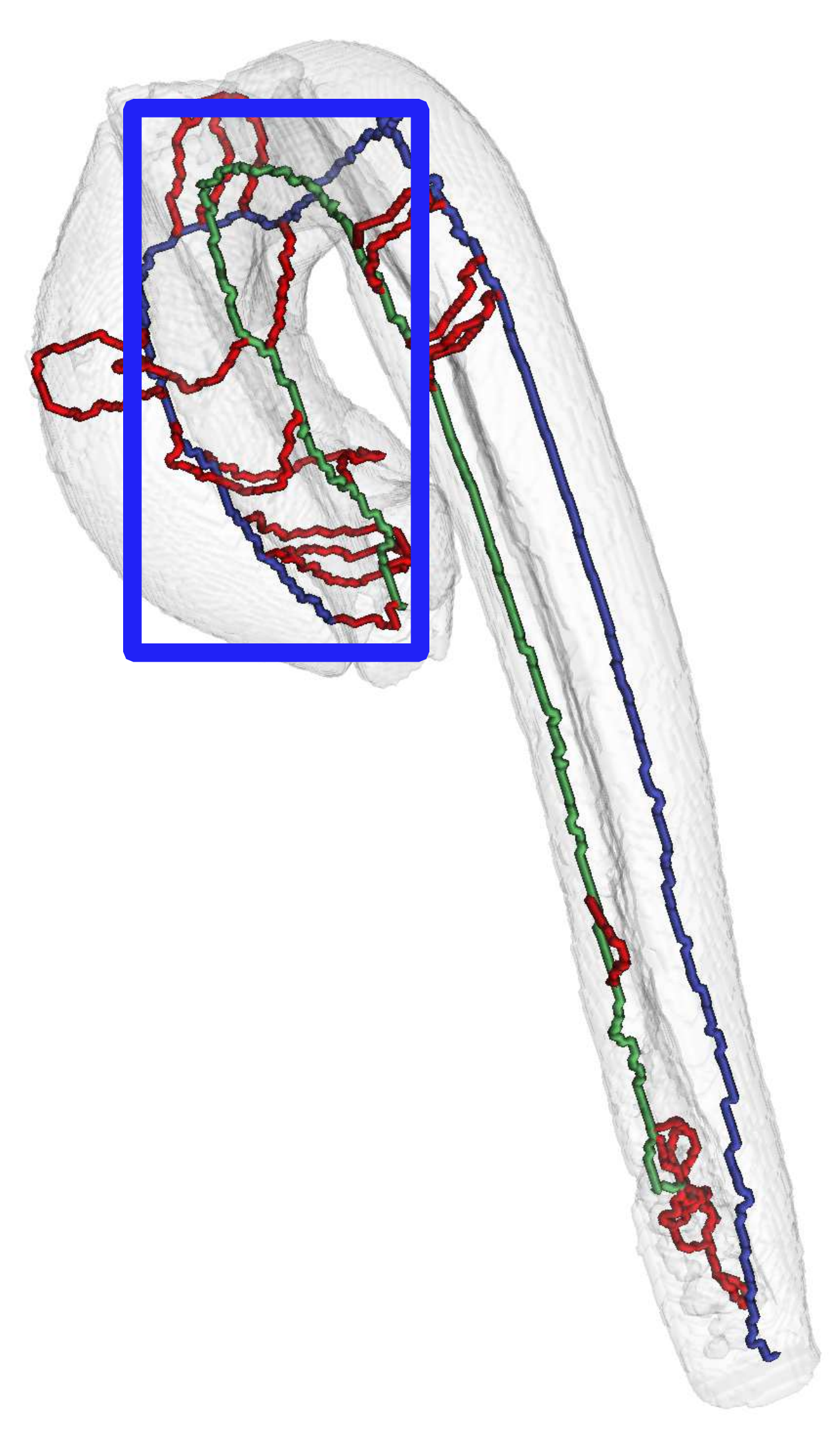} &
\includegraphics[width=2.8cm]{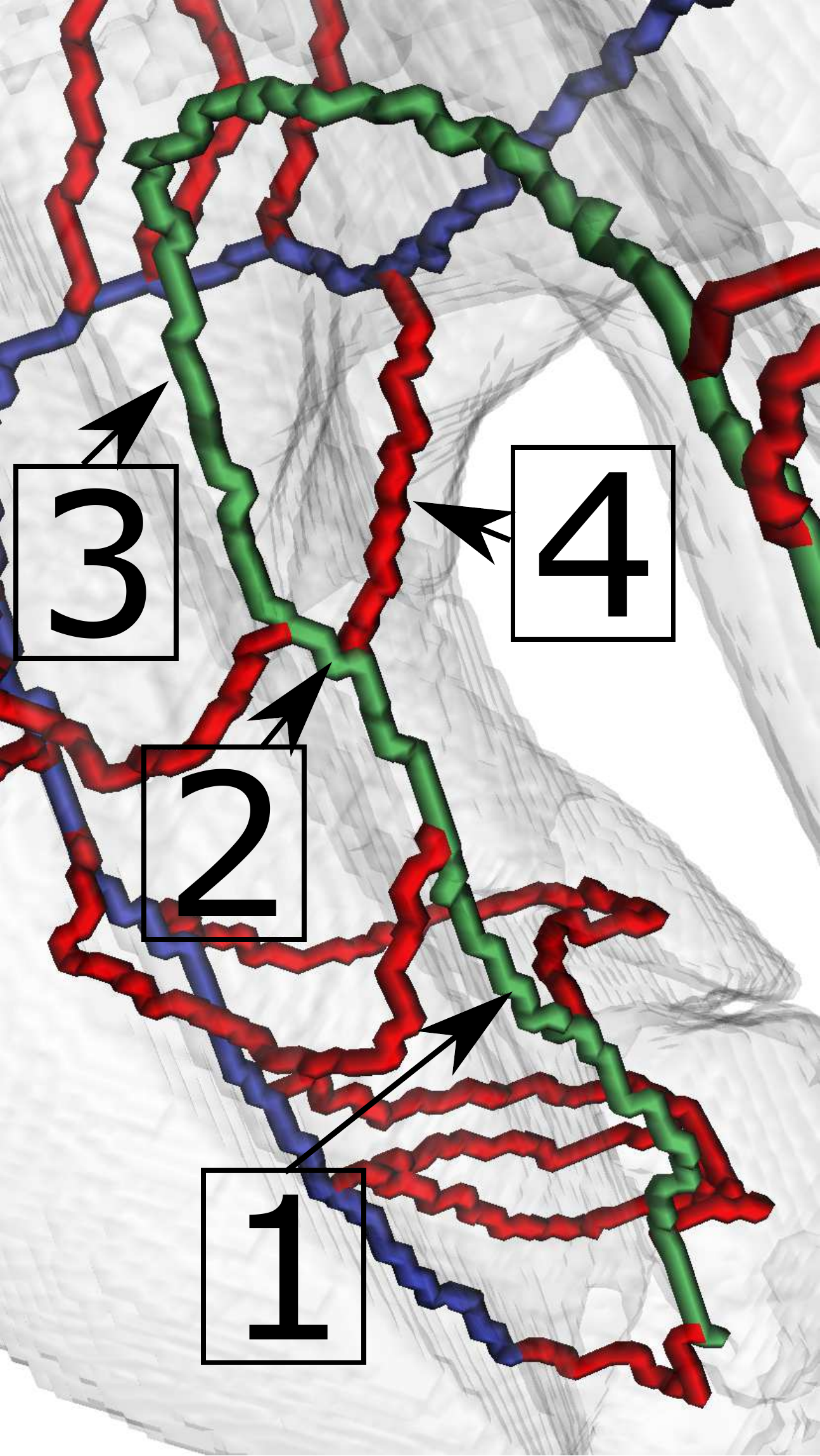} \\
(a) & (b) & (c) & (d) & (e) \\
\end{tabular}
\end{center}
\caption{(a) Initial image, (b) skeleton (algorithm proposed in Ref. \cite{Lohou2005}), (c) centerlines: green for a first lumen, blue for a second lumen, red for connections between lumens (corresponding to multiple entry tears), (d) another viewing orientation, (e) zoom (blue rectangle in (d)).}
\label{fig:fig5}
\end{figure}

One could then think of using the curvature along the skeleton during the processing of the shortest path in order to avoid going from one lumen to another. This way cannot be followed either: see Fig. \ref{fig:fig5} (e), the inspection of the curvature of the skeleton can make us follow the part of the skeleton between the two lumens that corresponds to a tear: one follows the green 1-2 segment in a lumen, at the point 2, one can either continue on the segment 2-3 (in the same lumen) or follow segment 2-4 (red) and reach the other lumen (whose main path is blue).

Suppose now that there is no passage from one lumen to another in the shortest path, \textit{i.e.}, we have obtained a well-identified centerline per lumen, we then carry out the conditional dilation of the centerline of each lumen. We consider that the conditional dilation is operated in such a way that one dilation iteration is made for one centerline, then for the other, and so on; the end of the algorithm is then obtained when there is no more point added (thus, there is no parameter of number of iterations to fix). 

\begin{figure}[t!]
\begin{center}
\begin{tabular}{ccccc}
\includegraphics[width=2.8cm]{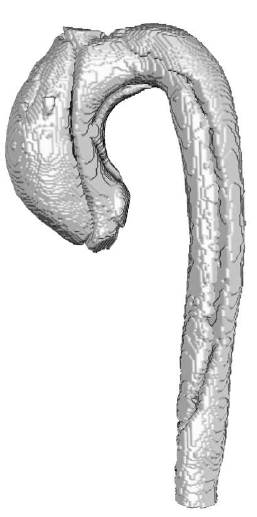} &
\includegraphics[width=2.8cm]{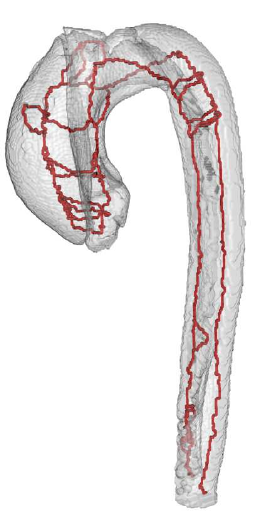} &
\includegraphics[width=2.8cm]{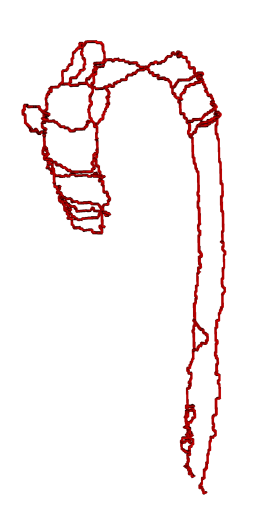} &
\includegraphics[width=2.8cm]{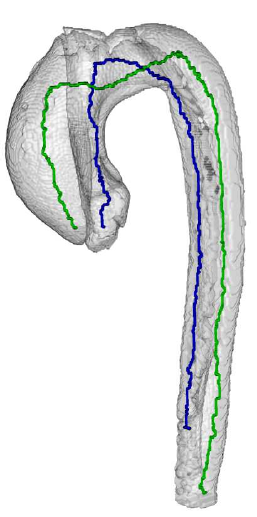} &
\includegraphics[width=2.8cm]{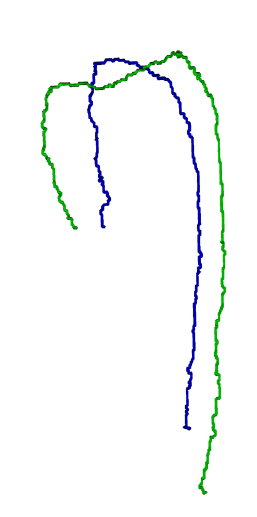} \\
(a) & (b) & (c) & (d) & (e) \\
\includegraphics[width=2.8cm]{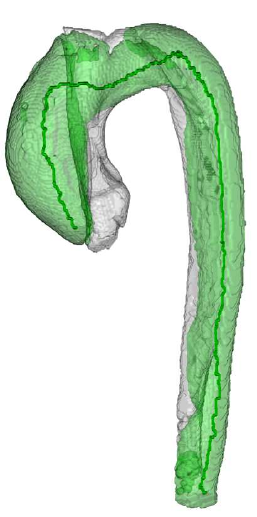} &
\includegraphics[width=2.8cm]{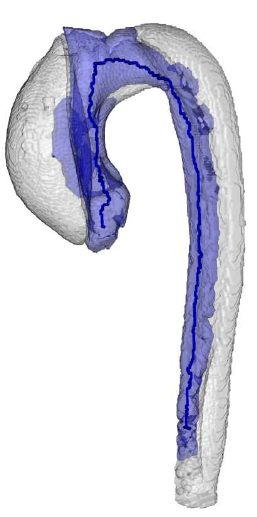} &
\includegraphics[width=2.8cm]{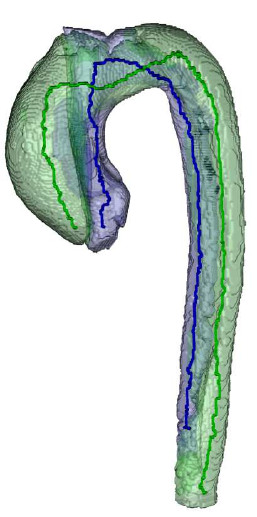} &
\includegraphics[width=2.8cm]{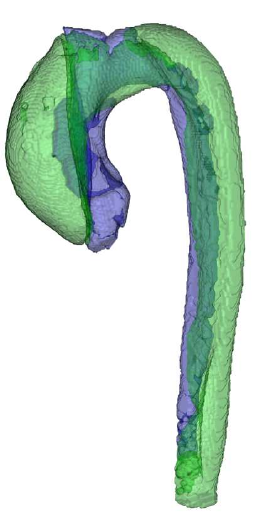} &
\includegraphics[width=2.8cm]{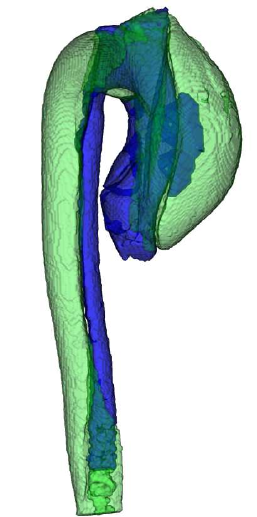} \\
(f) & (g) & (h) & (i) & (j)\\
\end{tabular}
\end{center}
\caption{(a) Initial image, (b) skeleton in red in the initial object in transparency, (c) skeleton only, (d) centerlines in the initial object in transparency, (e) centerlines only, (f) the dilation of the green centerline reaches the second lumen, (g) the dilation of the blue centerline in the second lumen, (h) union of the two dilations with the centerlines, (i-j) union of the dilations without centerlines (2 different views).}
\label{fig:fig6}
\end{figure}

Let us illustrate the result on the image in Fig. \ref{fig:fig6}. This example shows that the dilation of a centerline of a first lumen can reach the second one 
(Fig. \ref{fig:fig6} (f)): it is therefore not a way to follow either to disconnnect lumens.

\item By adapting a segmentation algorithm by deformable model.

A region growing segmentation may require a structuring element large enough to retrieve the lumens, but a large size of some tears would cause the propagation to pass from one lumen to another; a criterion based on gray level also does not allow the propagation to stop because of the near intensity of voxels in the neighborhood of tears. One study explored a fast-marching segmentation by adapting the velocity function to account for both the homogeneity of gray level in lumens and the edges of the lumens (gradient information); but segmentation results were not accurate enough \cite{Fetnaci2013}.

\end{itemize}

\subsubsection{Assessment of direct or indirect approaches}

Therefore, it seems difficult to semi-automatically retrieve the lumens separated, because of the size of the tear that can be large compared to the "size" of at least one lumen and also because gray level intensities of blood may be homogeneous around tears for the two lumens. In the next section, we have chosen to focus on the segmentation of connected lumens (as in our previous works) and then on their separation (indirect approach, as argued in Sect. \ref{sec:indirect_separation})).

\subsection{Our motivation}

The intimal tears of an aortic dissection are holes in the intimal wall. An algorithm for closing the holes in 3D objects has been proposed by exploiting Digital Topology operators \cite{Aktouf2002}. 
The principle of the algorithm is as follows: 
\begin{itemize} 
\item the initial object, made of $3$D points/voxels, is immersed in a filled bounding box; 
\item the image of the distance of the points relative to the initial object is computed, 
the chosen distance is relative to the object connectivity ($6$, $18$, or $26$ \cite{Kong1989}) 
and is a parameter of the algorithm; 
\item the points of the bounding box are arranged inside a hierarchical list, 
that is an array comprising as many elements as the maximum distance value 
and for which each index element $i$ stores a queue (First In First Out) of points of distance value $i$; 
\item an output image is created: the hierarchical list and the output image are initialized by the points of the outer layer of the bounding box; 
\item we examine the highest priority points (those furthest from the initial object) of the hierarchical list that do not belong to the initial object: 
if they are at a distance less than a certain threshold (second parameter of the algorithm) 
and if they do not check a certain property - see after -, 
they are removed from the hierarchical list and from the output image. 
The neighboring points (according to the connectivity of the object) of the points examined in the hierarchical list are added both in the hierarchical list 
according to their distance and in the output image if they were not already there. 
The priority is adjusted if necessary (if neighboring points added during the current iteration are further than the distance value being examined). 
This process is repeated until all points have been examined.
\end{itemize}

The property that the points must verify to be kept is to separate the complementary of object being built (output image) into two parts. It is decided in the local neighborhood of the examined point thanks to the topological numbers \cite{Bertrand1994}. 
Therefore, points constituting a surface filling a hole of the initial object will be preserved, more precisely, only surfaces whose size are greater than the second parameter are kept. Also, the output image is then composed of the points of the initial image and the points of surfaces filling holes of a minimal size. 

\begin{figure}[t!]
\begin{center}
\begin{tabular}{cc}
\includegraphics[width=4cm]{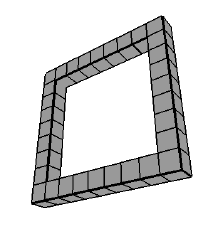} & \includegraphics[width=4cm]{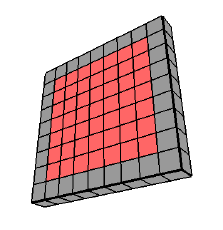} \\
(a) & (b)\\
\end{tabular}
\end{center}
\caption{(a) Initial object, (b) surface, in red, closing the hole of the initial object.}
\label{fig:fig7}
\end{figure}

In Fig. \ref{fig:fig7}, we illustrate the result on a synthetic object (the distance used is $d_{26}$ \cite{Kong1989}, and the size of the holes to be closed is less than or equal to $4$).

\begin{figure}[t!]
\begin{center}
\begin{tabular}{cccccc}
\includegraphics[height=5.5cm]{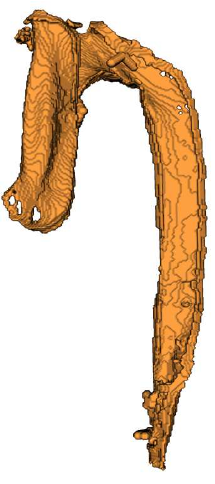} &
\includegraphics[height=5.5cm]{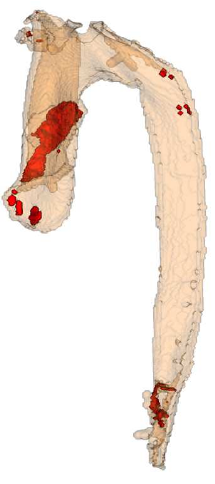} &
\includegraphics[height=5.5cm]{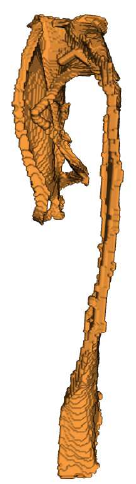} &
\includegraphics[height=5.5cm]{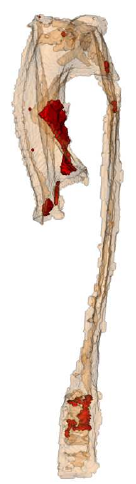} &
\includegraphics[height=5.5cm]{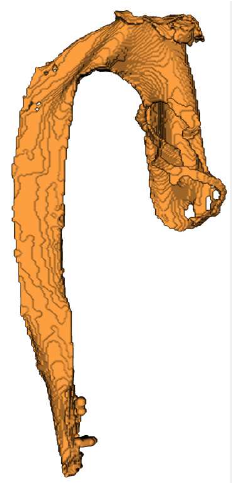} &
\includegraphics[height=5.5cm]{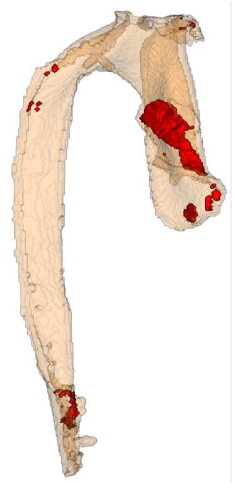} \\
(a) & (b) & (c) & (d) & (e) & (f)\\
\end{tabular}
\end{center}
\caption{(a, c, e) Initial image of the flap ($3$ different orientations), (b, d, f) filled flap (in semi-transparency) for the same $3$ orientations, surfaces filling holes are drawn in red.}
\label{fig:fig8}
\end{figure}

Fig. \ref{fig:fig8} shows the result of the holes closing algorithm applied to an image of flap: surfaces that fill holes are drawn in red. There are several tears: each of these doors corresponds to a connection between the two lumens either because there is a real entry tear or because the segmentation is such that the two lumens have joined at this area 
(according to the precision of the segmentation process). We will now focus on the original use of filling surfaces in order to disconnect the lumens obtained by segmentation.

\section{Our contribution}\label{sec:our_contribution}

By setting as zero the second parameter of the holes closing algorithm (minimal size of surfaces filling holes), 
we will obtain all the surfaces that correspond to connections (of any size) 
between lumens (more precisely, either real tears or joined area in the segmentation results, as explained before). 
By simply performing a set difference between the image of connected lumens and the image of surfaces filling holes, 
then we obtain the image of disconnected lumens. 
At our knowledge, this is the first study to use the surfaces filling holes to disconnect several parts of a 3D digital object. 

We may consider the flap as being $n$-connected. 
Thus, we also consider the surfaces filling holes with $n$-adjacency. 
Therefore, the flap/intimal tears set constitutes a $n$-connected object which separates the lumens. 
To respect the discrete analog of Jordan's theorem \cite{Kong1989}, 
we must consider the lumens to be $\overline{n}$-connected to be separated by the $n$-connected flap-intimal tears set. 
In the following, we illustrate results with $n=6$ and $\overline{n}=26$, in other words, 
with the $6$-connectivity for flap/tears set and $26$-connectivity for lumens.

\section{Method}\label{sec:method}

Once the DICOM image of aortic dissection is loaded, we first delineate a tubular area of interest around the aorta \cite{Lubniewski2012}, we segment the blood with a region growing algorithm, we obtain the image of connected lumens, the small vessels leaving the aorta are not retained because they do not belong to the area of interest (if necessary we can filter them by a series of erosions then dilations). Then we dilate $n$ times then erode $n$ times the image of connected lumens, we remove from the latter the image of the connected lumens and we get the image of the volume between the two lumens, that corresponds to the flap \cite{Lohou2013a}. 
We then apply the holes closing algorithm on this flap image to fill it, then by difference between the last image of the filled flap and the image of flap, we get the image of the surfaces filling holes \cite{Lohou2011}. 
Then, as proposed in this paper, we remove these surfaces from the image of the connected lumens and thus obtain the image of the disconnected lumens (cf. Graphical Abstract).

Several remarks: 
\begin{itemize}
\item We segment the blood in the lumens: each connection between these lumens corresponds either to one or several real tears, 
or to areas where lumens are joining according to the precision of the segmentation process as written before. 
In all cases, the disconnection well operates on these two cases.
\item We assume that the succession of dilations then erosions lead to encircle each connection in the computed image of the flap, otherwise, a surface will not be obtained by the holes-closing algorithm and this connection will not be removed. 
This may occurs when an intimal tear is located on the edge of a lumen: in this case, for example, we may expand the flap by computing the shortest path between two points manually defined at the edge of the non-closed contour around the connection; this path must be included in the complementary of connected lumens (to be sure that the surface that will  fill such a contour will well cut the whole corresponding connection).
\end{itemize}

\section{Results}\label{sec:results}

\subsection{3D Visualization}

We consider an aortic dissection image (initial size: $475 \times 512 \times 512$, crop size: $200 \times 200 \times 350$). 
Subsequently, we use Mitk \cite{Wolf2005} software for visualization. In Fig. \ref{fig:fig9} and Fig. \ref{fig:fig10}, 
we show the successive images (connected lumens, flap, tears) that lead to the image of disconnected lumens (Fig. \ref{fig:fig10} (b-c)).

We end with the image presenting the flap, the intimal tears and each separate lumen, in other words the cartography of an aortic dissection, Fig. \ref{fig:fig10} (f). 
To our knowledge, that leads to the first cartography of an aortic dissection by a semi-automatical process, 
\textit{i.e.}, without a manual delineation of the intimal tears and lumens. 
We highlight that it is quite impossible to manually delineate a $3$D hole/a tear 
in successive 2D cross-sections of a $3$D CTA image.

\subsection{Visualization of results by sets of successive slices}

We now illustrate the process of lumens separation with a focus around on a little intimal tear, on a series of ten successive slices. Fig. \ref{fig:fig11}, are represented the contours (in white) for the connected lumens, the flap, the intimal tear, the result of the separation of the lumens by subtraction of the intimal tear; magnified images showing the satisfying disconnection are also given.

\begin{landscape}
\begin{figure}[t!]
\begin{center}
\begin{tabular}{ccc}
\includegraphics[height=6.5cm]{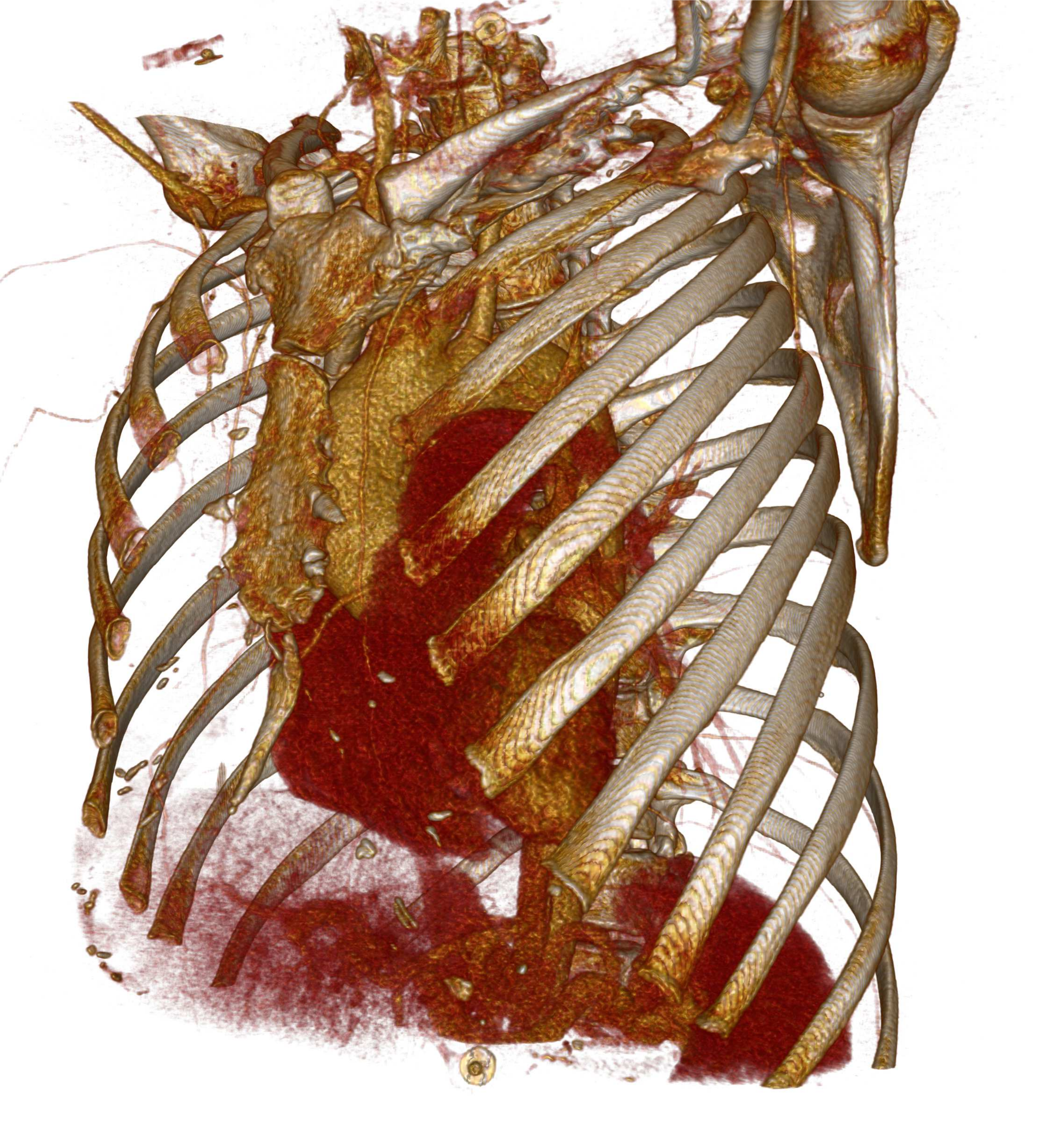} & 
\includegraphics[height=6.5cm]{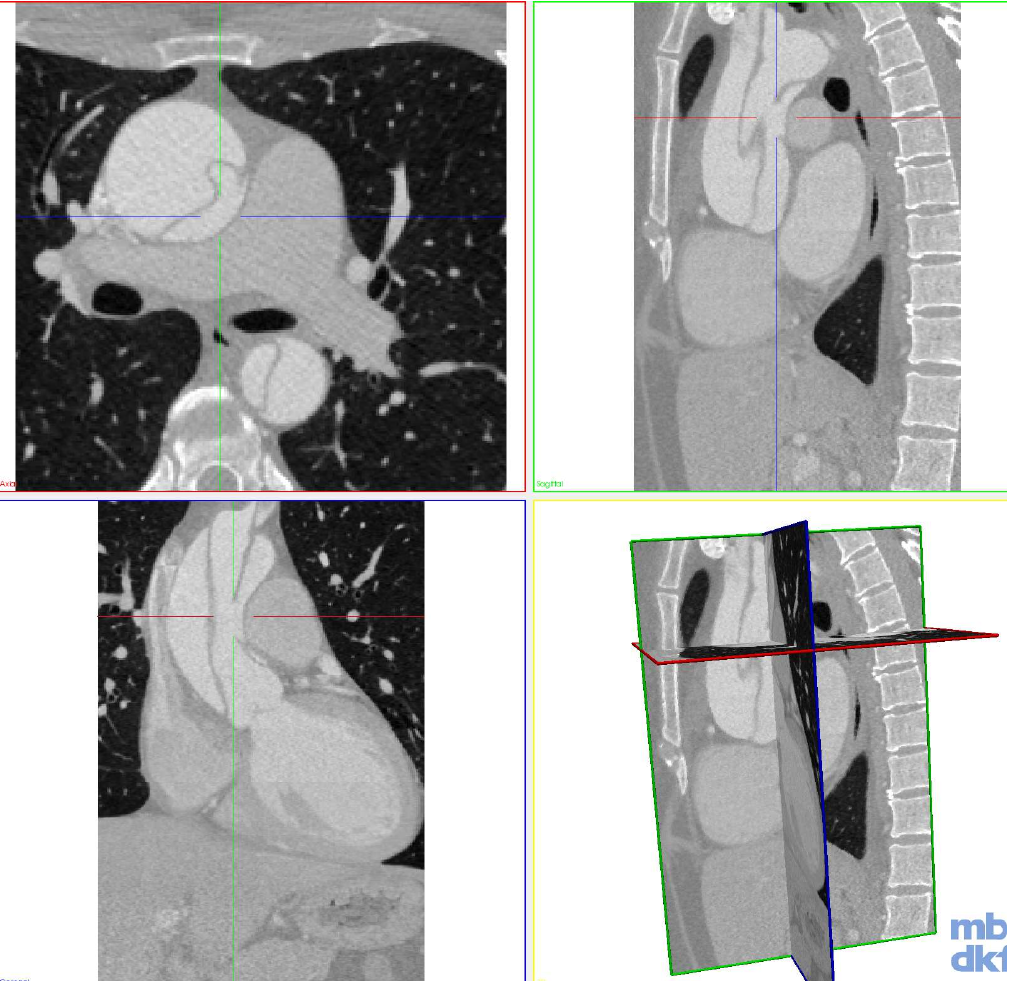} &
\includegraphics[height=6.5cm]{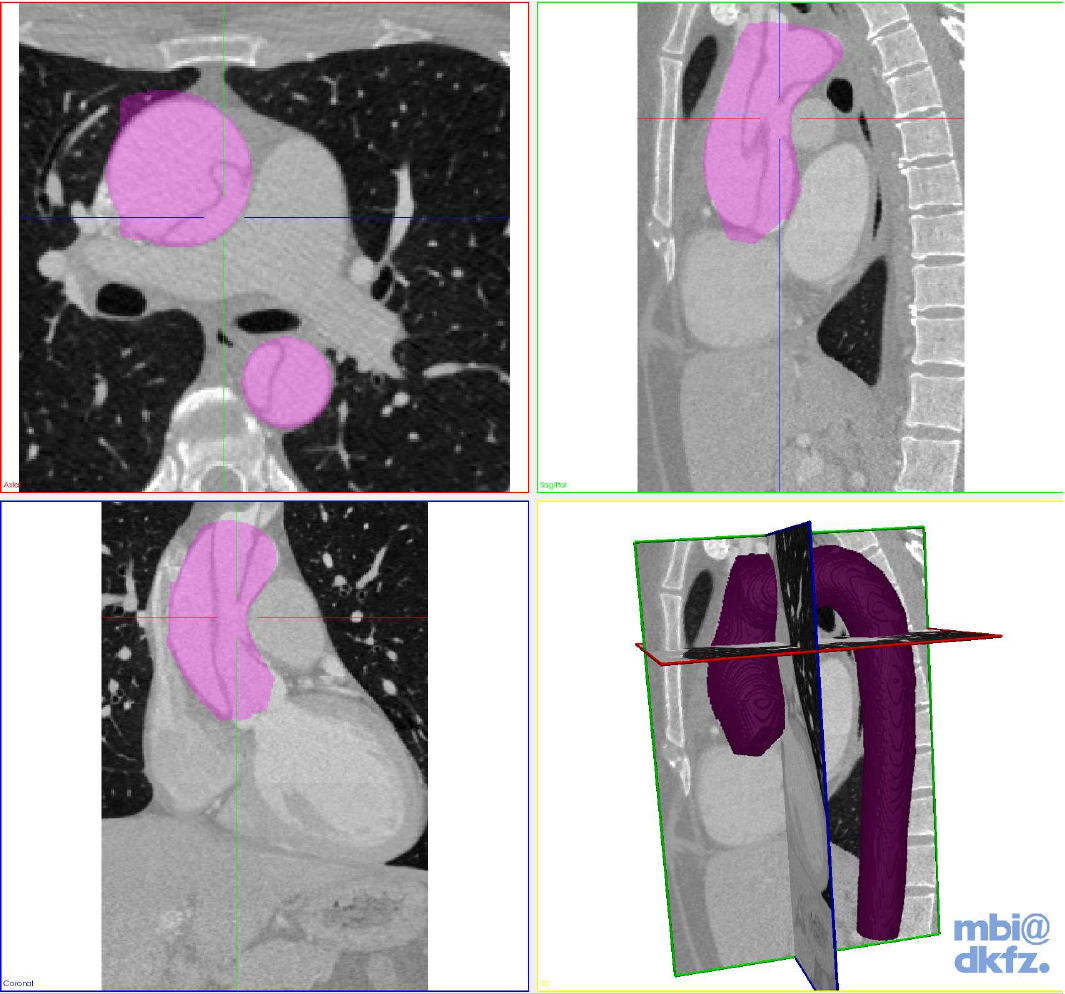} \\ 
(a) & (b) & (c)\\
\includegraphics[height=6.5cm]{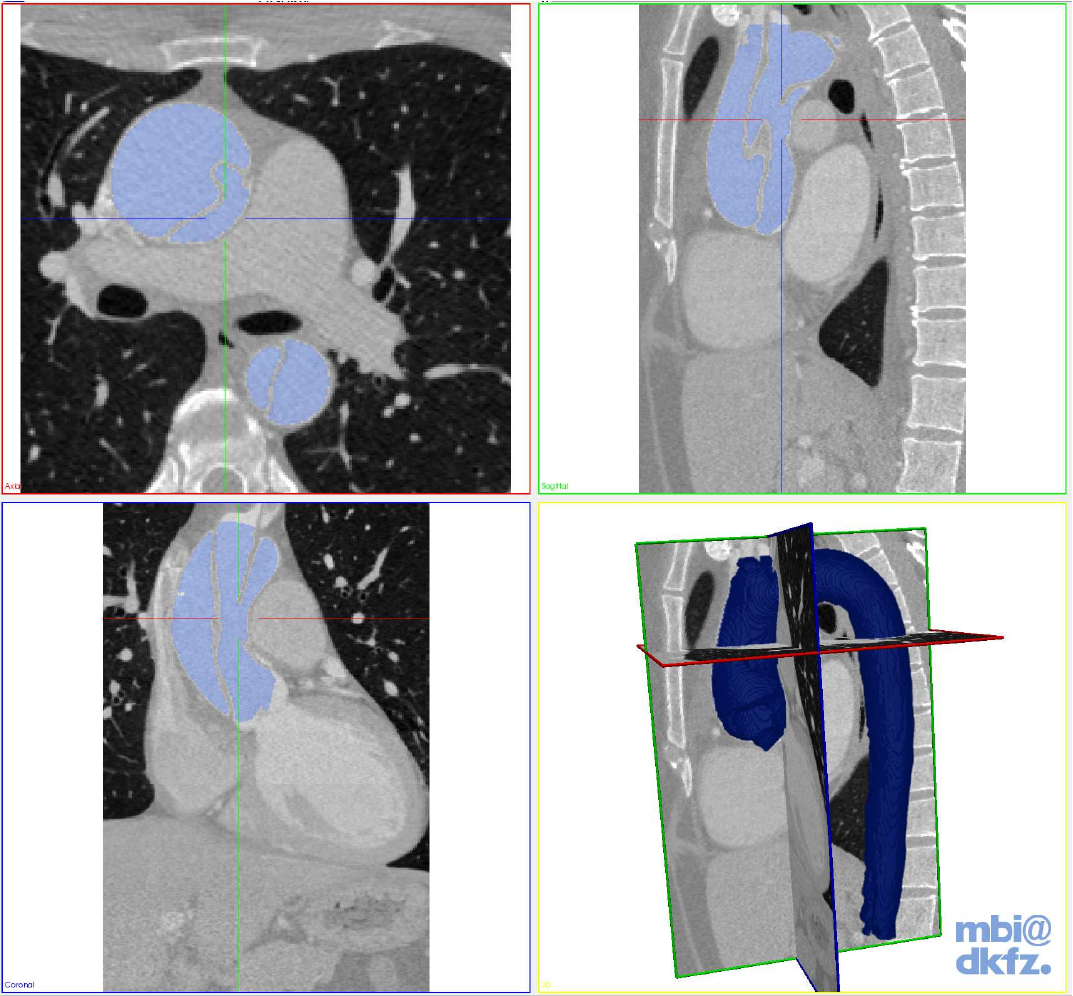} &
\includegraphics[height=6.5cm]{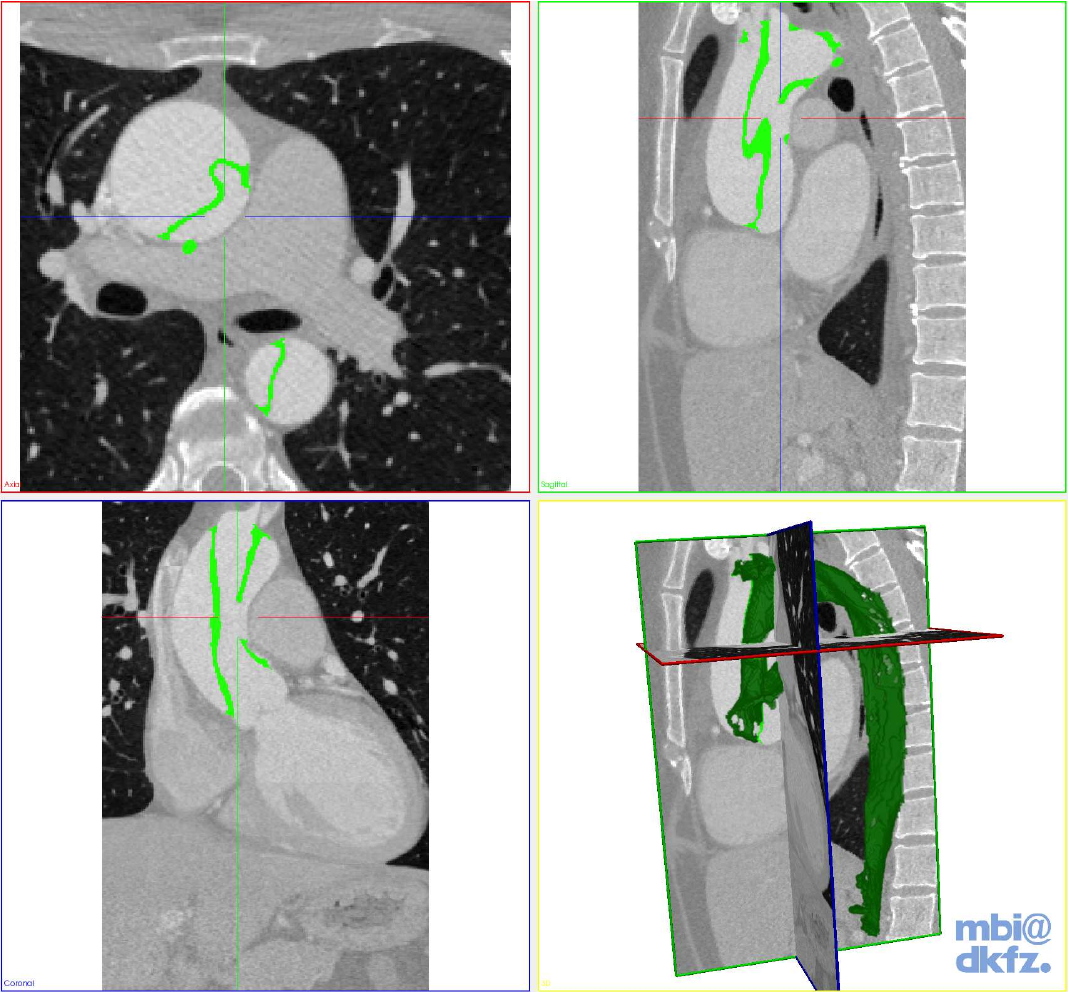} & 
\includegraphics[height=6.5cm]{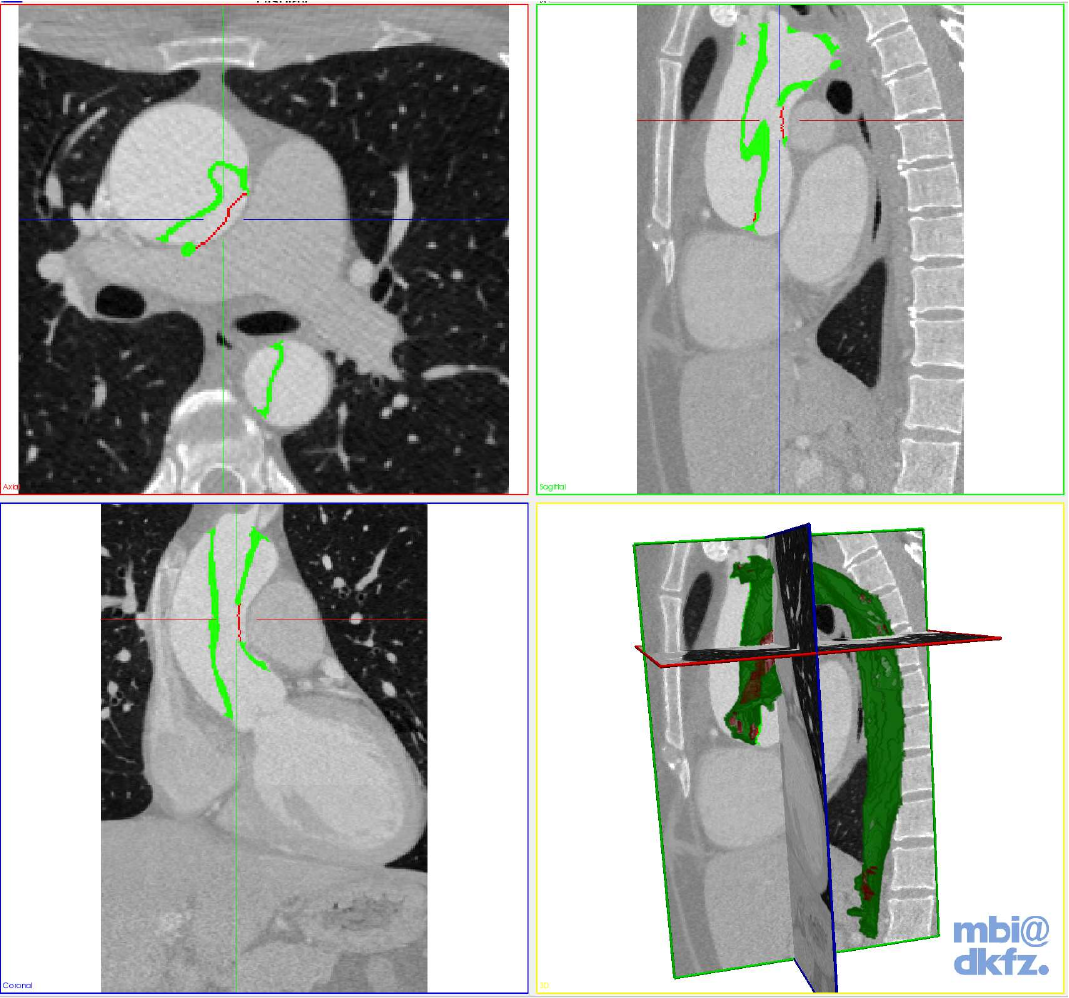} \\
(d) & (e) & (f) \\
\end{tabular}
\end{center}
\caption{(a) Volume rendering of a 3D CTA image of aortic dissection, (b) rendering in 2D sections and 3D of the CTA, (c) tubular area of interest - crop (in purple), (d) result of segmentation of connected lumens, (e) flap (in green), (f) filling surfaces (in red) of $3$D holes of the flap (in green). The cutting planes intersect themselves at the main intimal tear of this dissection.}
\label{fig:fig9}
\end{figure}
\end{landscape}

\begin{landscape}
\begin{figure}[t!]
\begin{center}
\begin{tabular}{ccc}
\includegraphics[height=6.5cm]{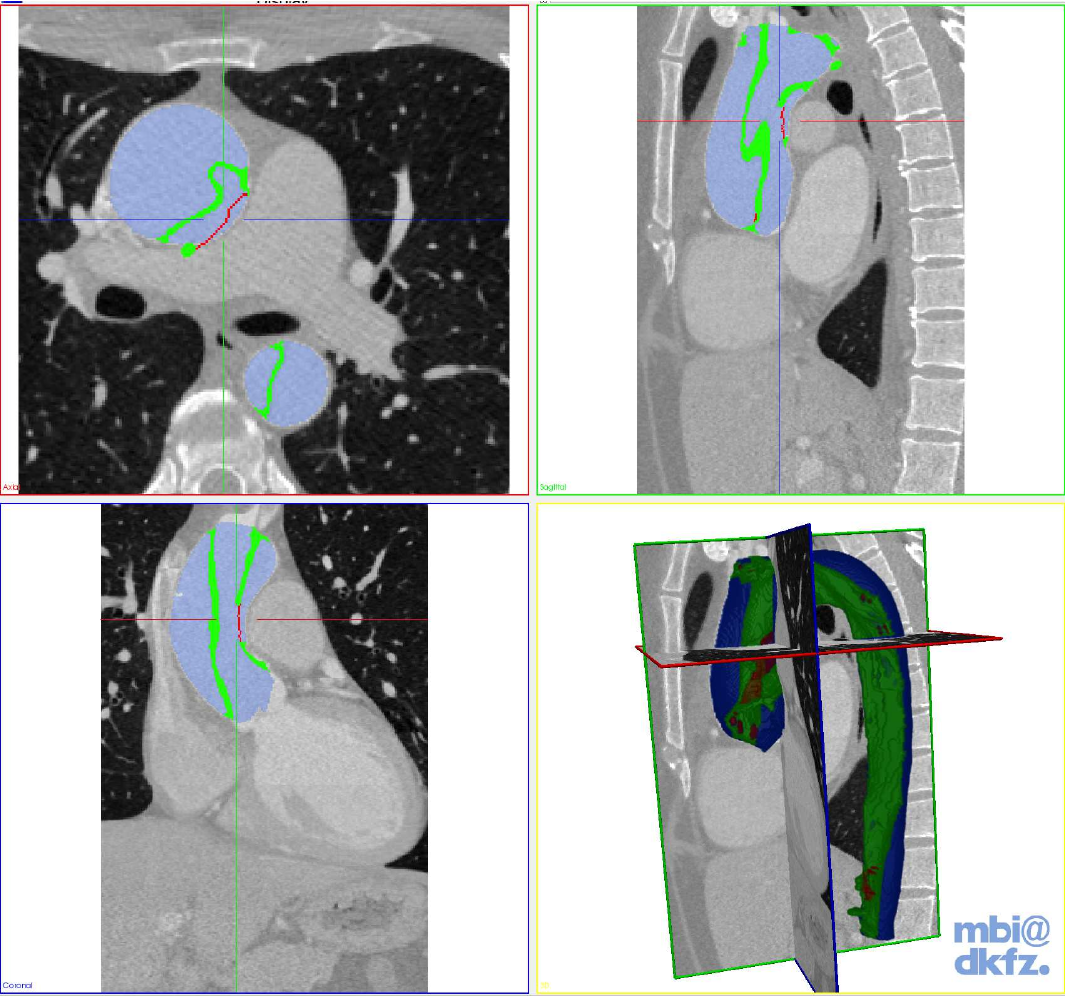} & 
\includegraphics[height=6.5cm]{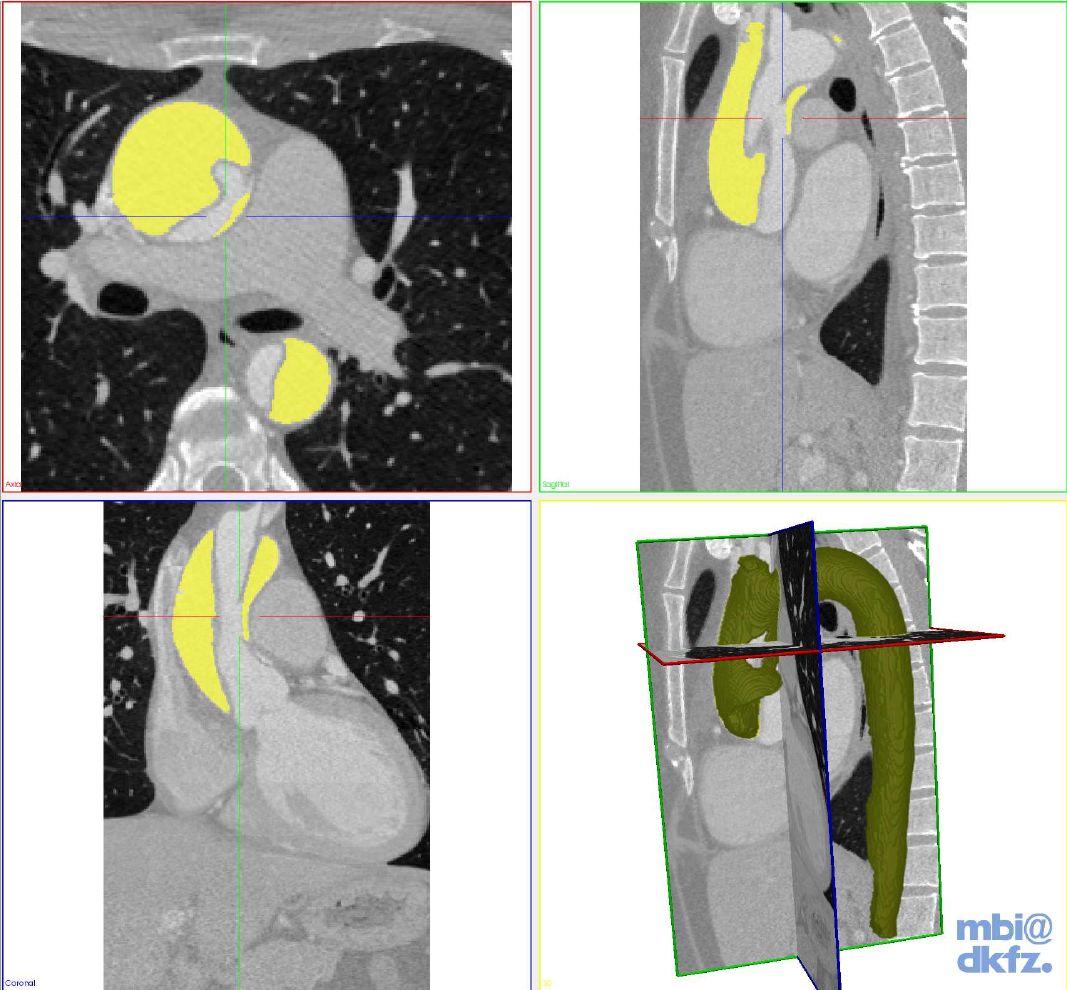} &
\includegraphics[height=6.5cm]{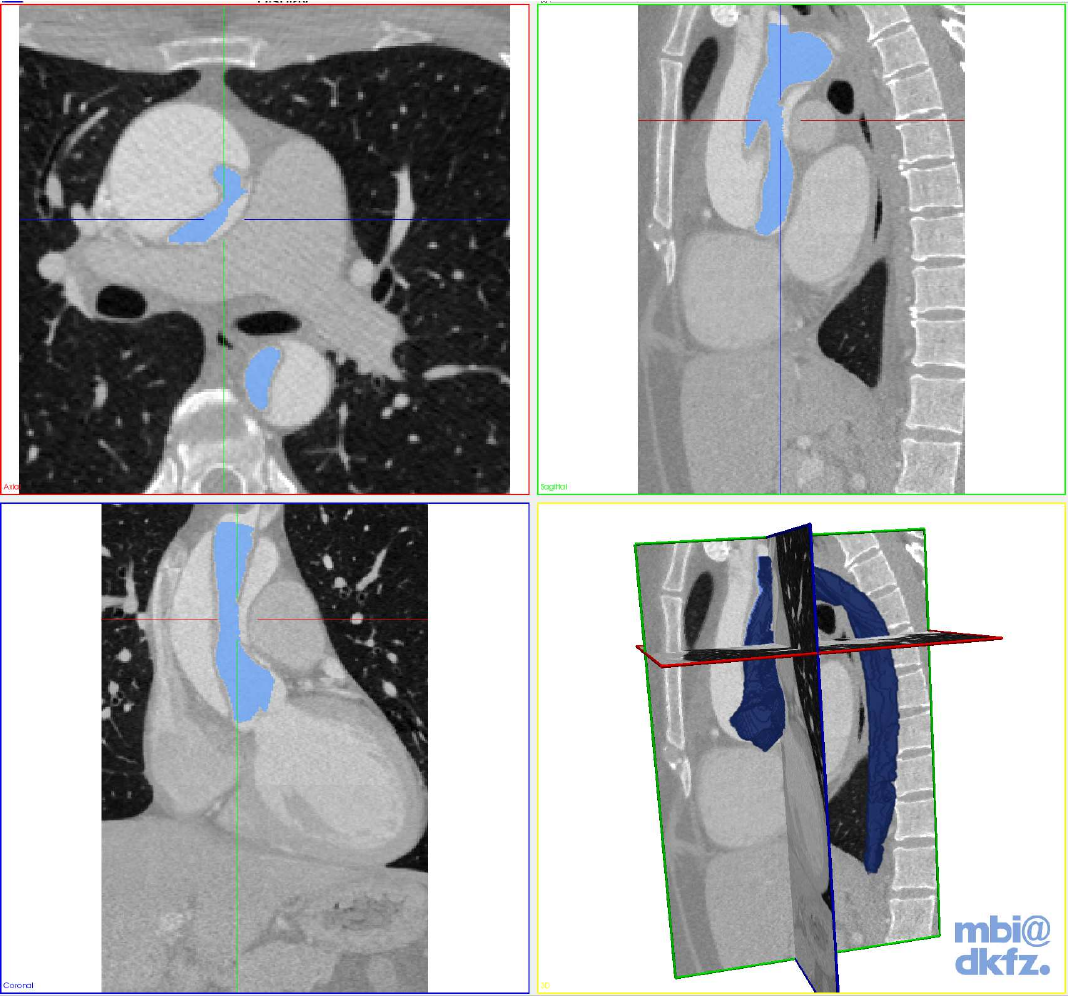} \\ 
(a) & (b) & (c)\\
\includegraphics[height=6.5cm]{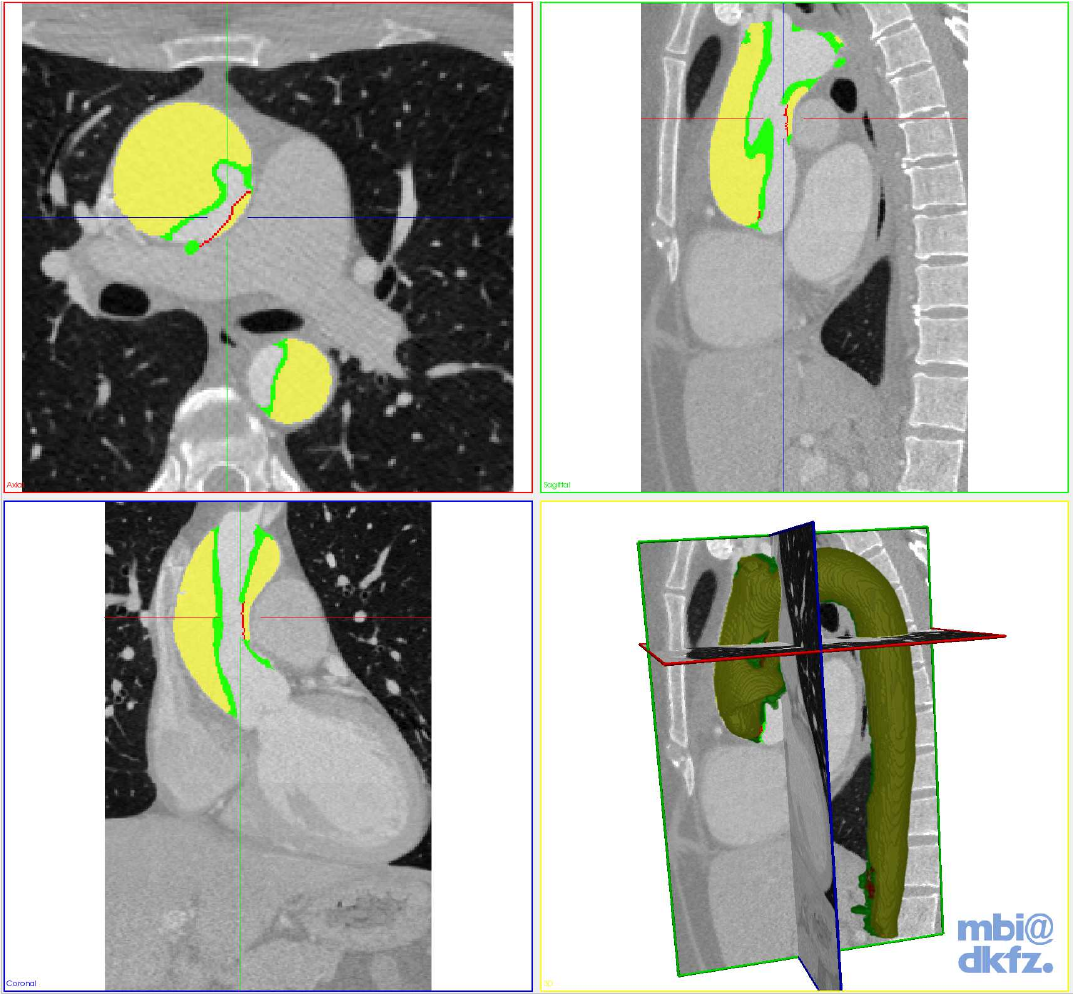} &
\includegraphics[height=6.5cm]{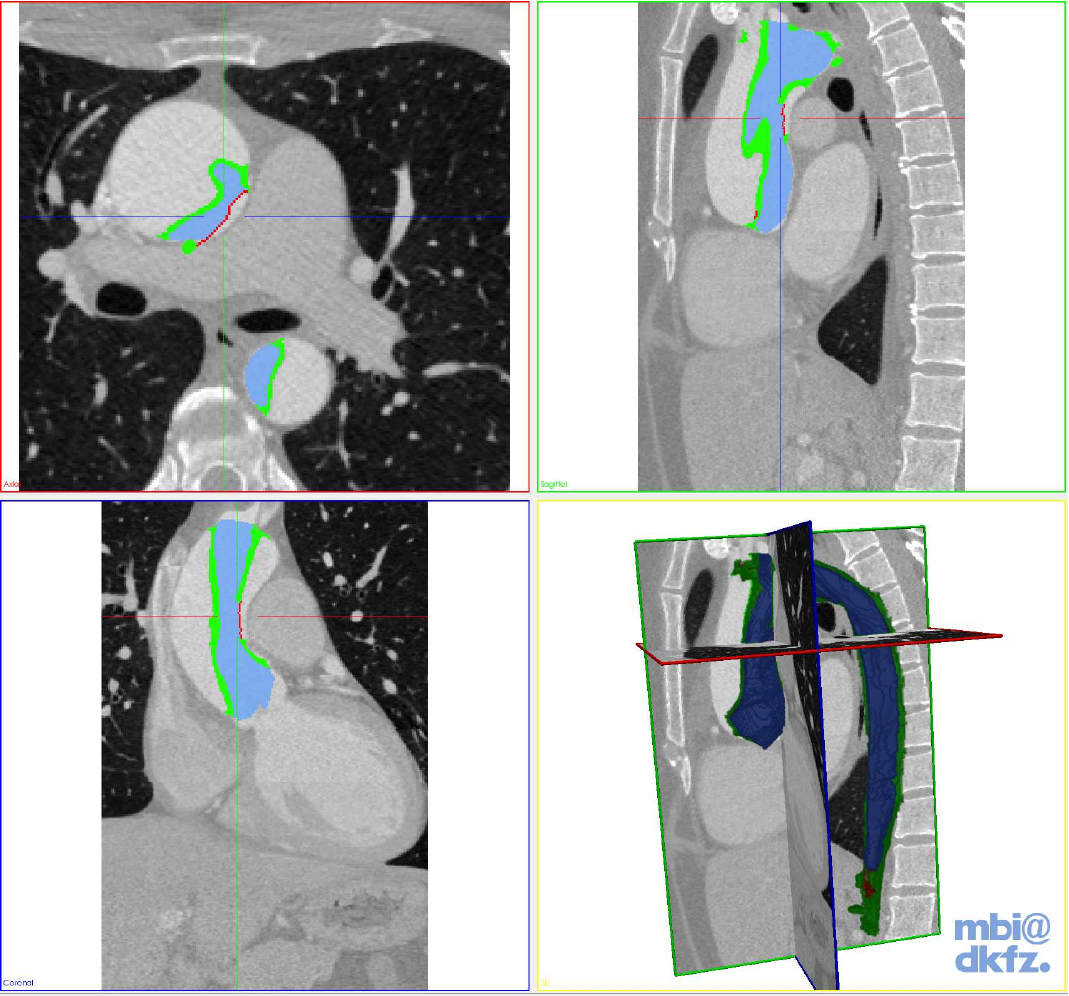} & 
\includegraphics[height=6.5cm]{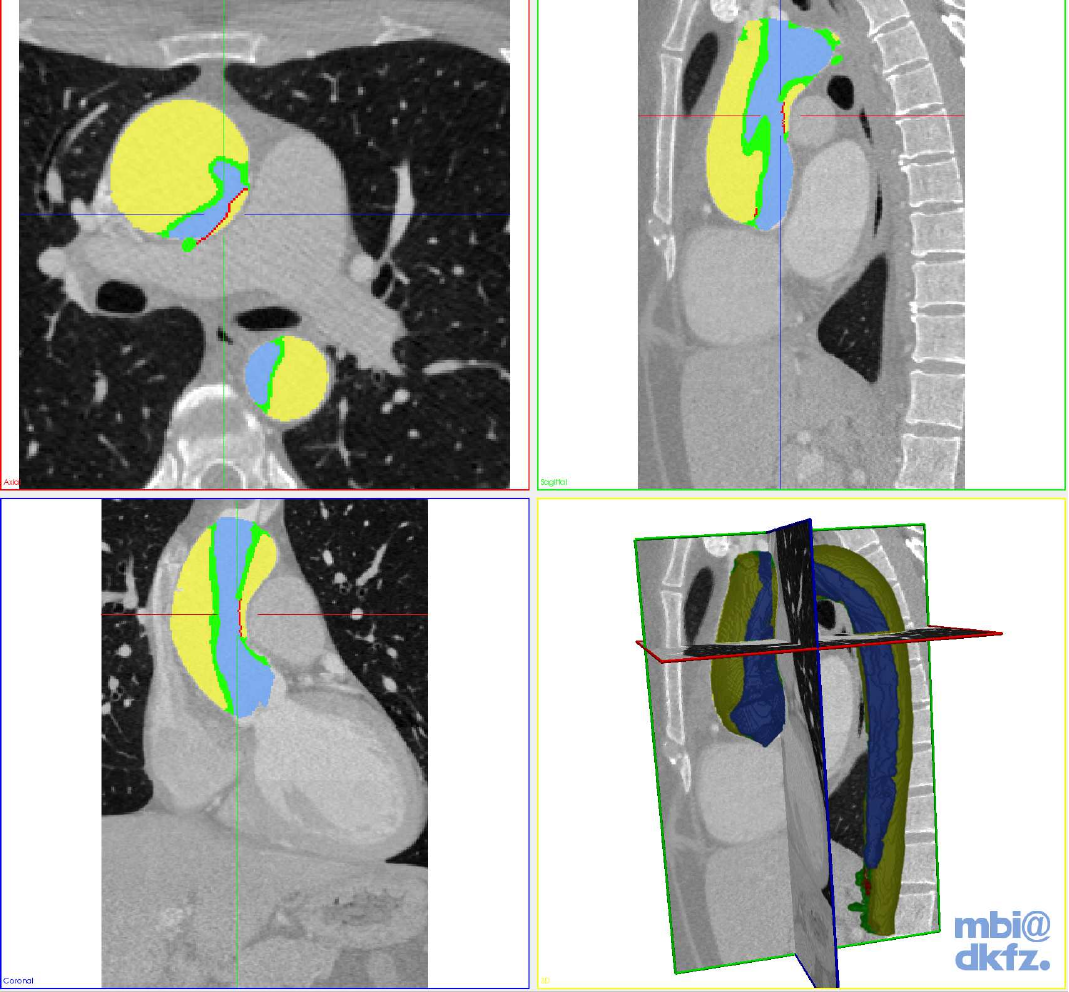} \\
(d) & (e) & (f) \\
\end{tabular}
\end{center}
\caption{(a) Filling surfaces (in red) in connected lumens (in blue) and in the flap (in green). After the disconnection: (b) first lumen (in yellow), (c) second lumen (in blue), (d) first lumen with flap and intimal tears, (e) second lumen with flap and tears, (f) cartography of the aortic dissection: flap, intimal tears and the two separated lumens.}
\label{fig:fig10}
\end{figure}
\end{landscape}

\begin{landscape}
\begin{figure}[t!]
\begin{center}
\begin{tabular}{cccccccccccc}
\raisebox{1.2cm}{(a)} & \includegraphics[width=1.8cm]{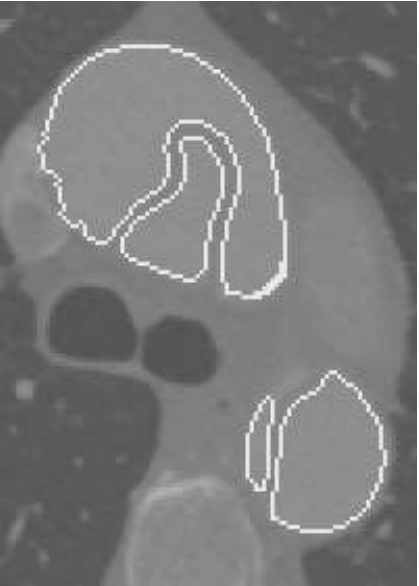}& 
\includegraphics[width=1.8cm]{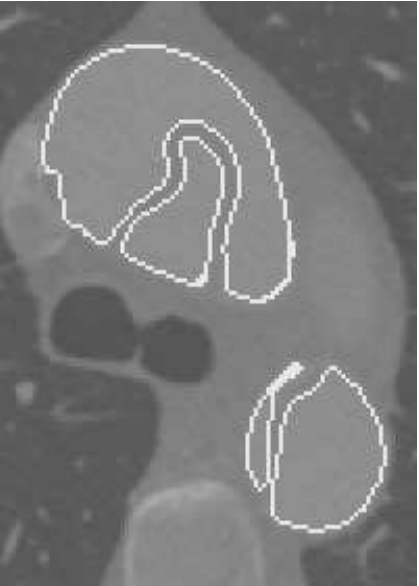}& 
\includegraphics[width=1.8cm]{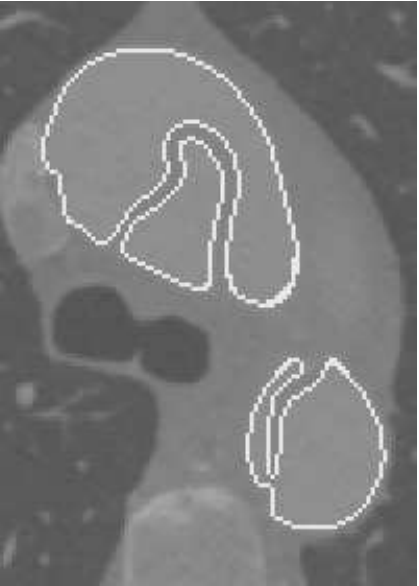}& 
\includegraphics[width=1.8cm]{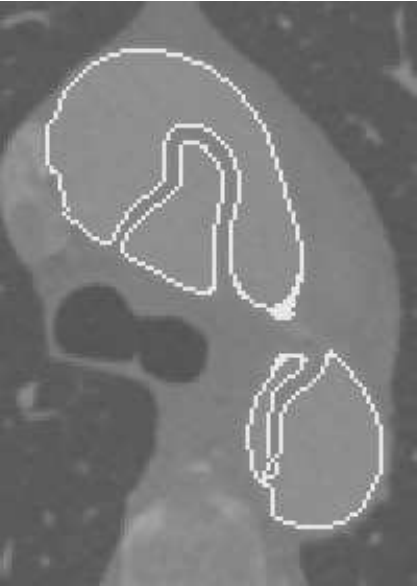}& 
\includegraphics[width=1.8cm]{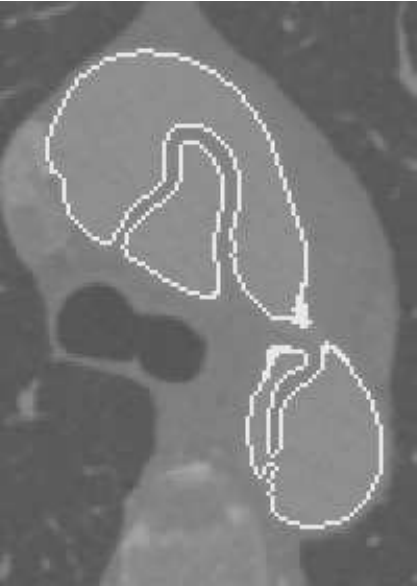}& 
\includegraphics[width=1.8cm]{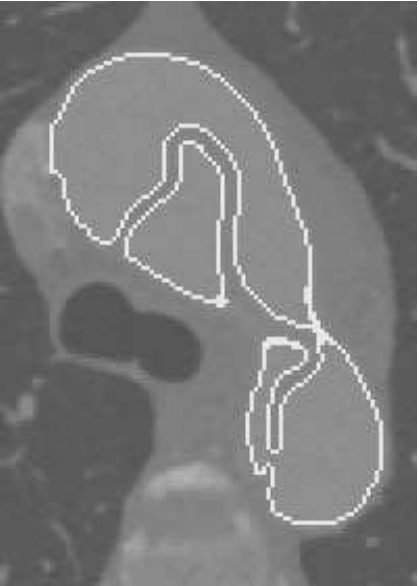}& 
\includegraphics[width=1.8cm]{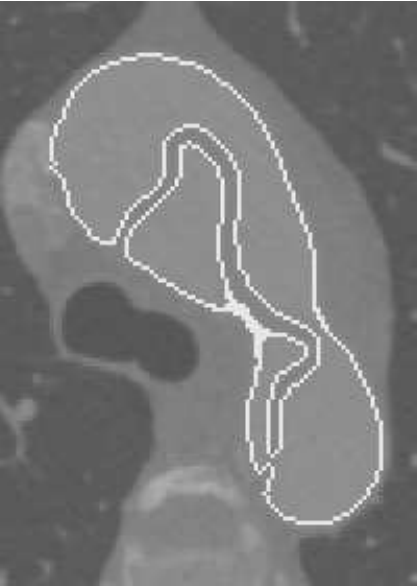}& 
\includegraphics[width=1.8cm]{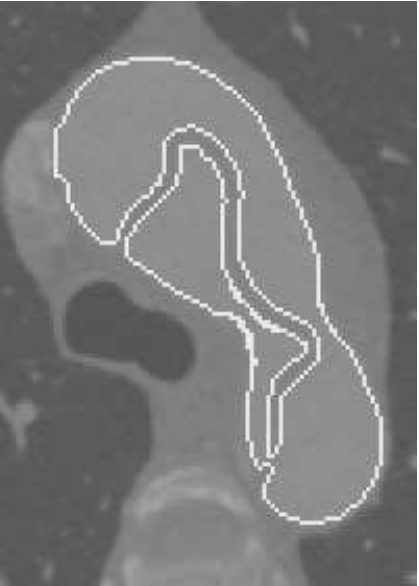}& 
\includegraphics[width=1.8cm]{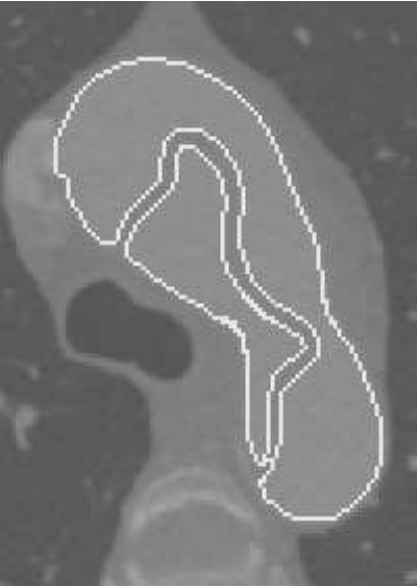}& 
\includegraphics[width=1.8cm]{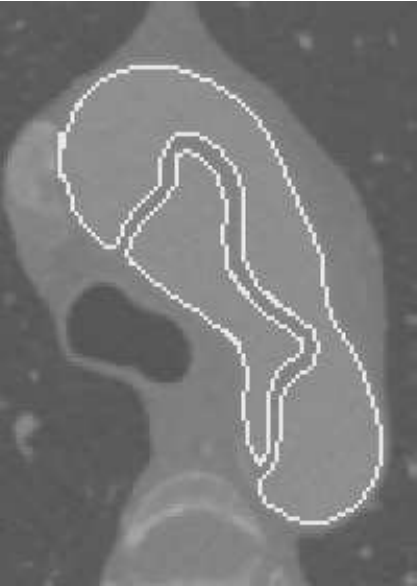}\\
\raisebox{1.2cm}{(b)} & \includegraphics[width=1.8cm]{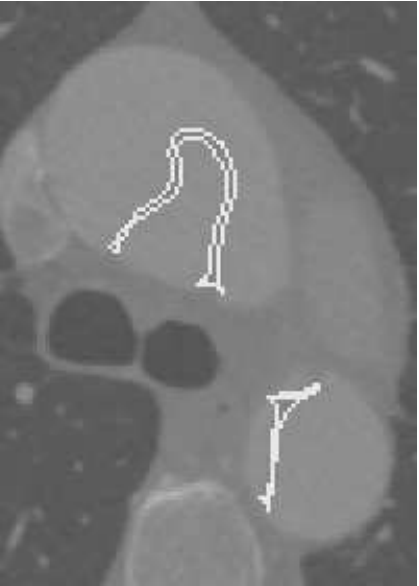}& 
\includegraphics[width=1.8cm]{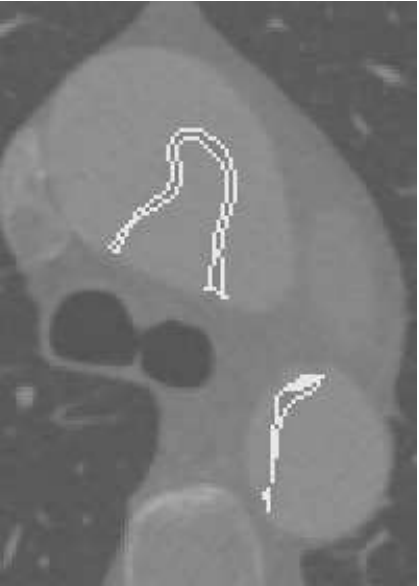}& 
\includegraphics[width=1.8cm]{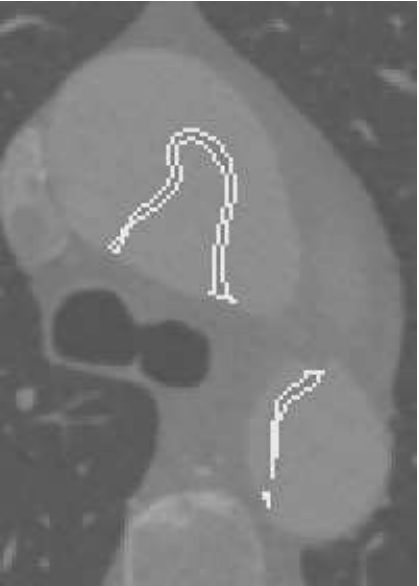}& 
\includegraphics[width=1.8cm]{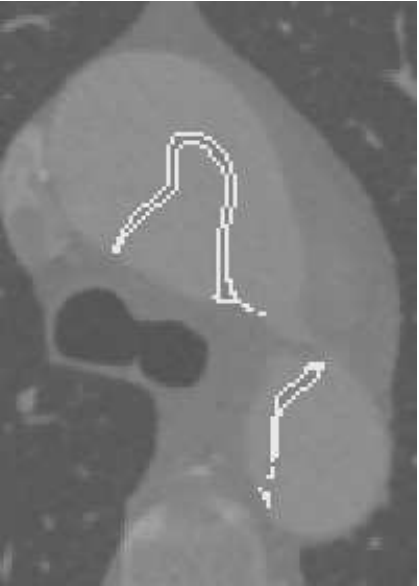}& 
\includegraphics[width=1.8cm]{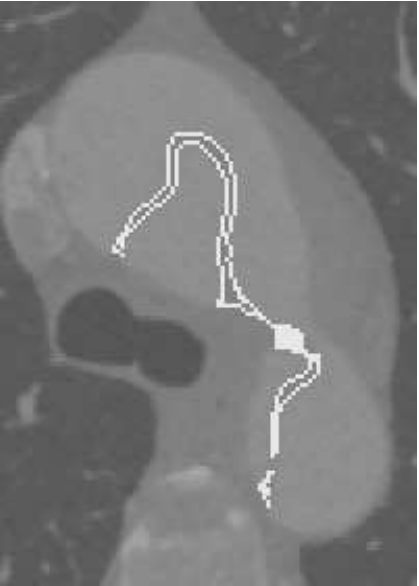}& 
\includegraphics[width=1.8cm]{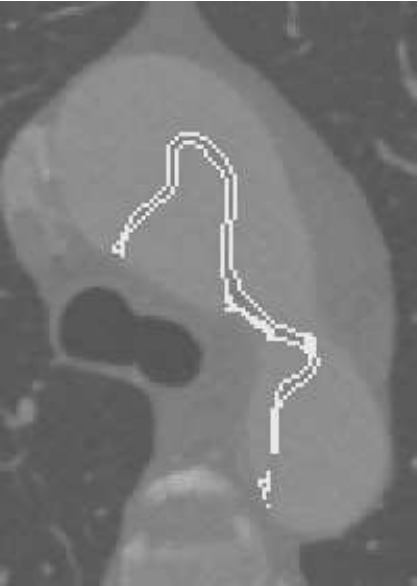}& 
\includegraphics[width=1.8cm]{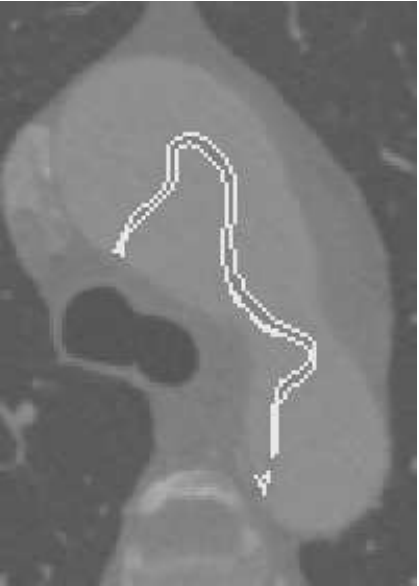}& 
\includegraphics[width=1.8cm]{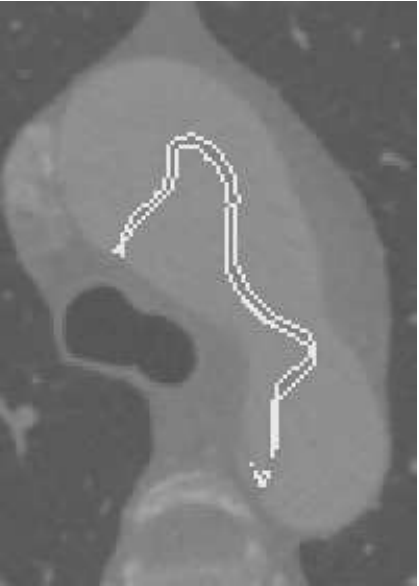}& 
\includegraphics[width=1.8cm]{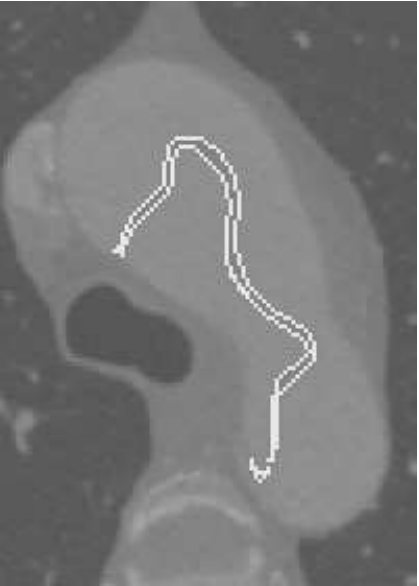}& 
\includegraphics[width=1.8cm]{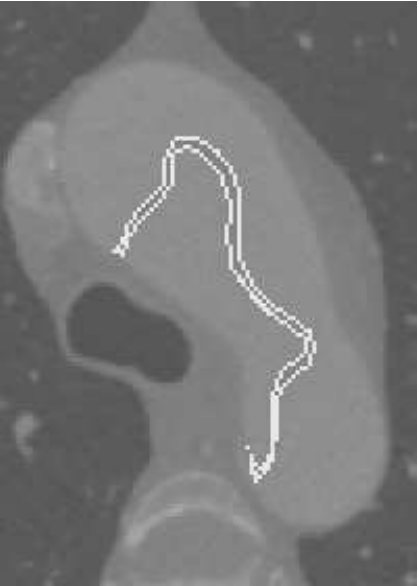}\\
\raisebox{1.2cm}{(c)} & \includegraphics[width=1.8cm]{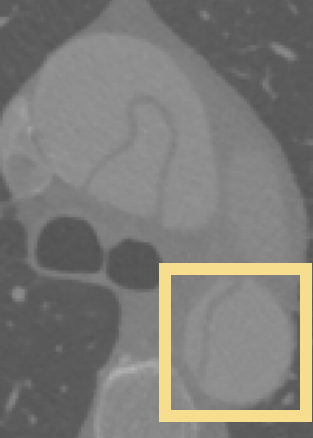}& 
\includegraphics[width=1.8cm]{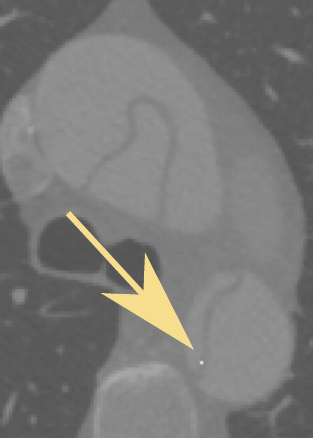}& 
\includegraphics[width=1.8cm]{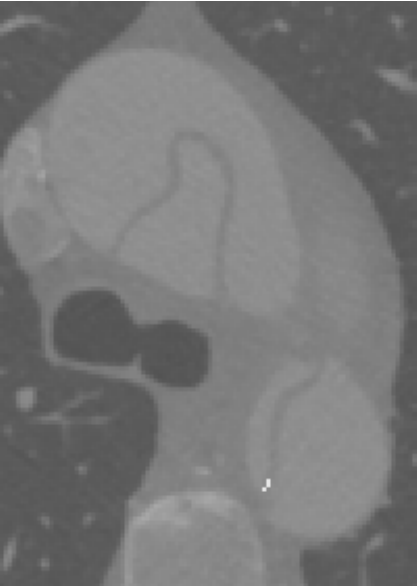}& 
\includegraphics[width=1.8cm]{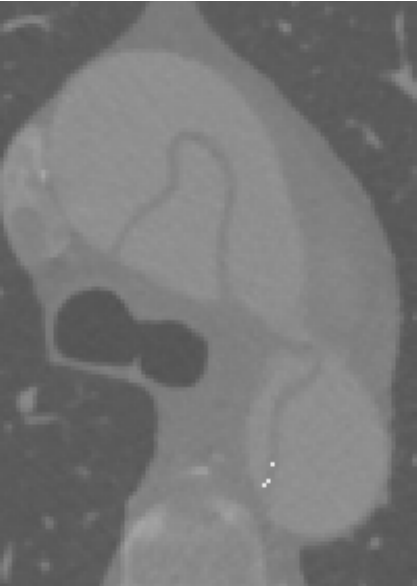}& 
\includegraphics[width=1.8cm]{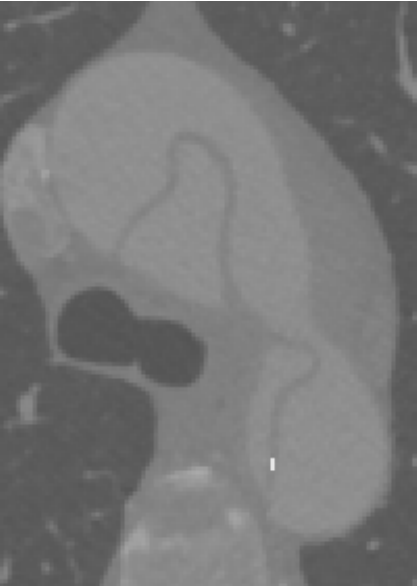}& 
\includegraphics[width=1.8cm]{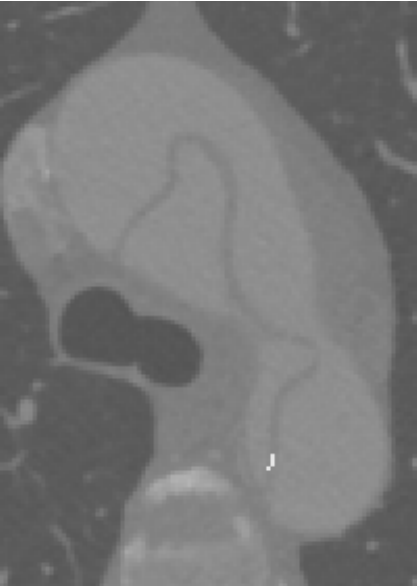}& 
\includegraphics[width=1.8cm]{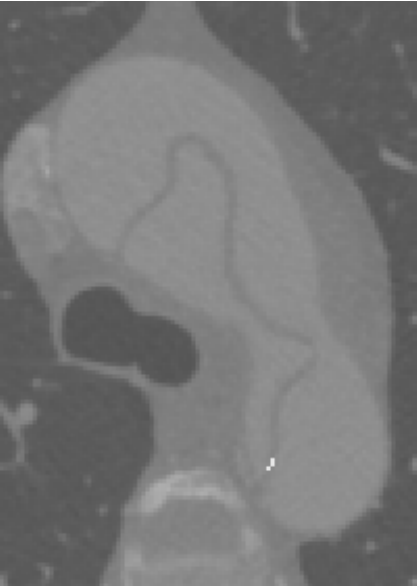}& 
\includegraphics[width=1.8cm]{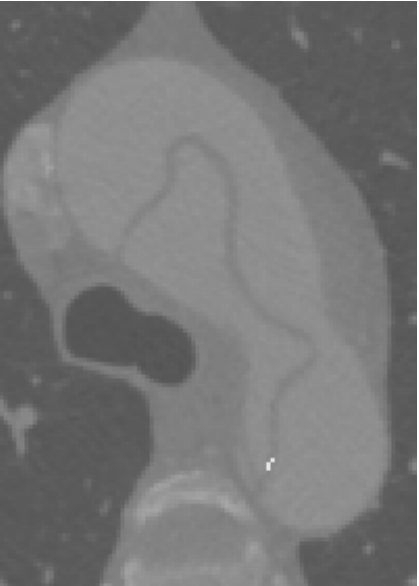}& 
\includegraphics[width=1.8cm]{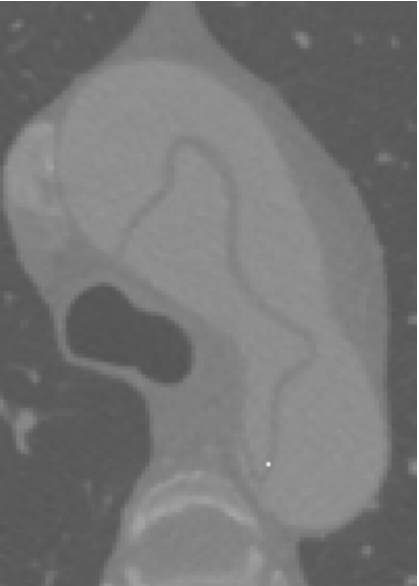}& 
\includegraphics[width=1.8cm]{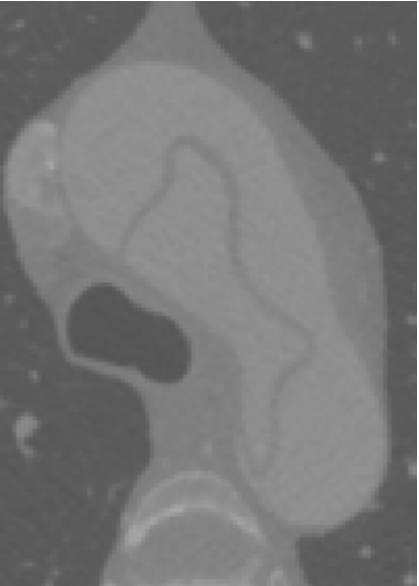}\\
\raisebox{1.2cm}{(d)} & \includegraphics[width=1.8cm]{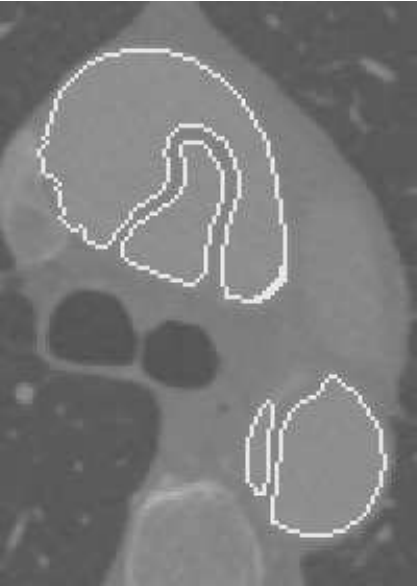}& 
\includegraphics[width=1.8cm]{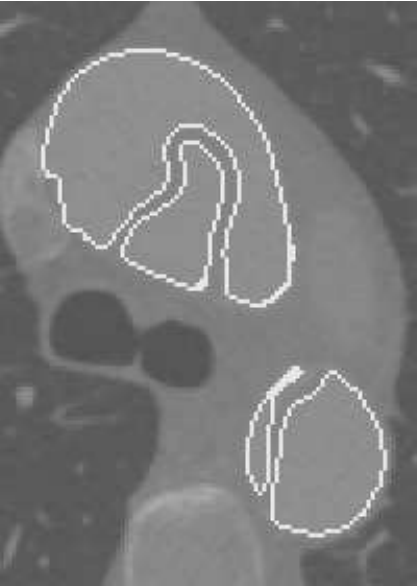}& 
\includegraphics[width=1.8cm]{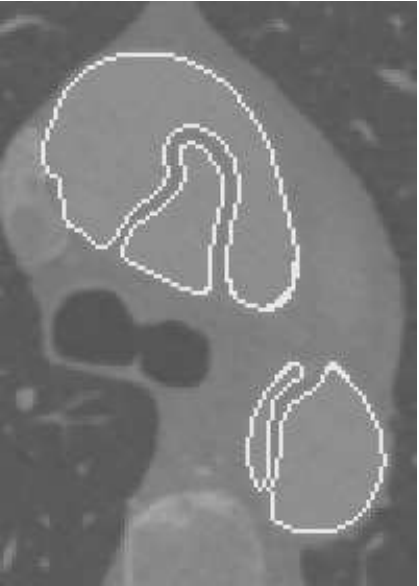}& 
\includegraphics[width=1.8cm]{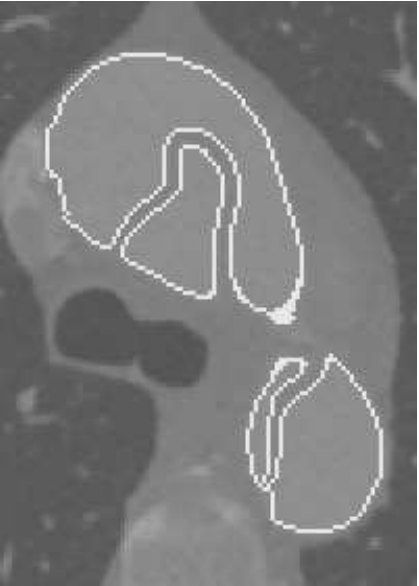}& 
\includegraphics[width=1.8cm]{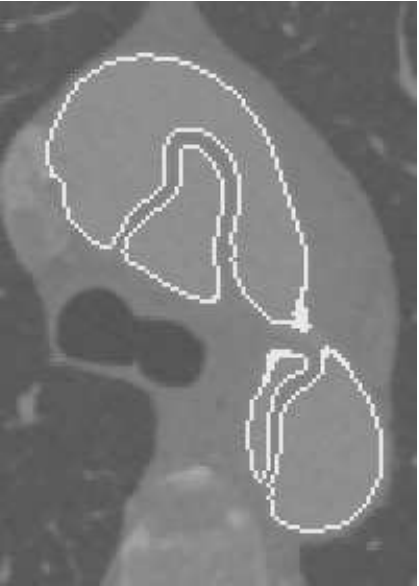}& 
\includegraphics[width=1.8cm]{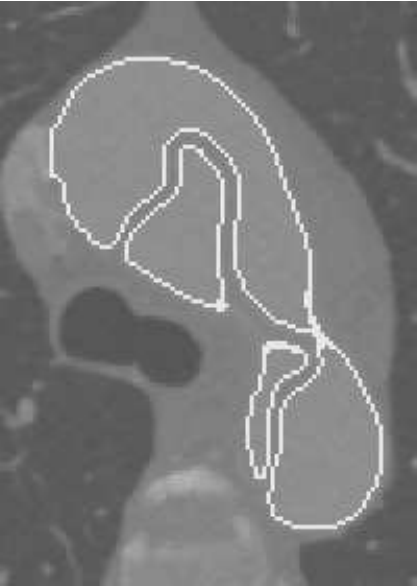}& 
\includegraphics[width=1.8cm]{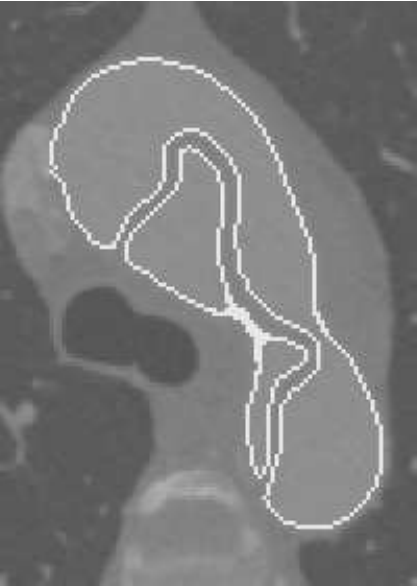}& 
\includegraphics[width=1.8cm]{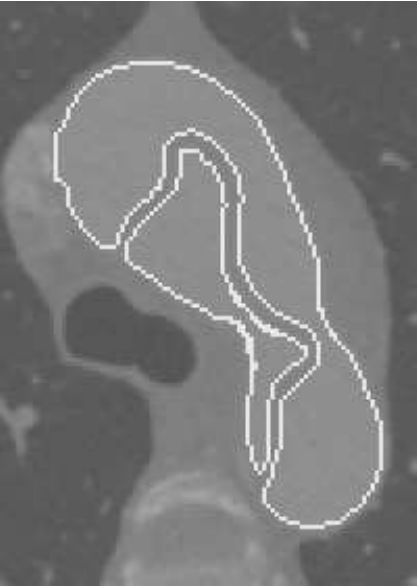}& 
\includegraphics[width=1.8cm]{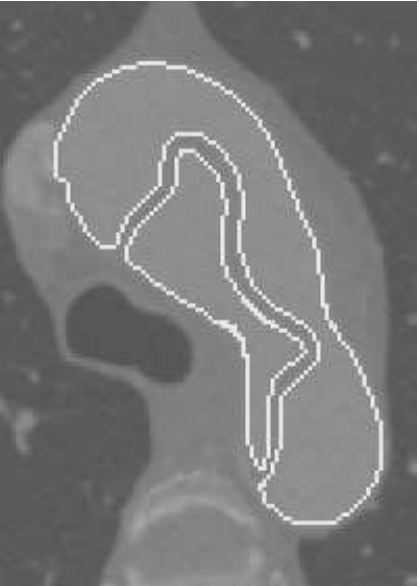}& 
\includegraphics[width=1.8cm]{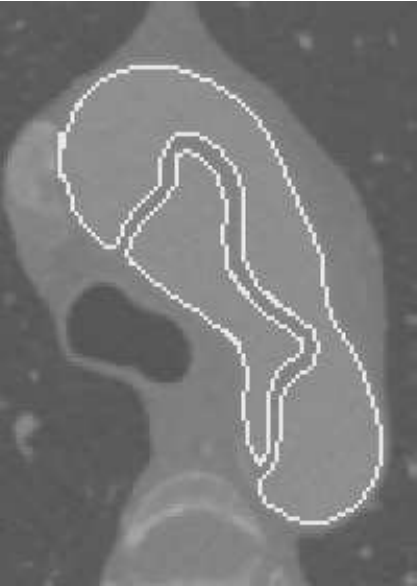}\\
\raisebox{0.9cm}{(e)} & \includegraphics[width=1.8cm]{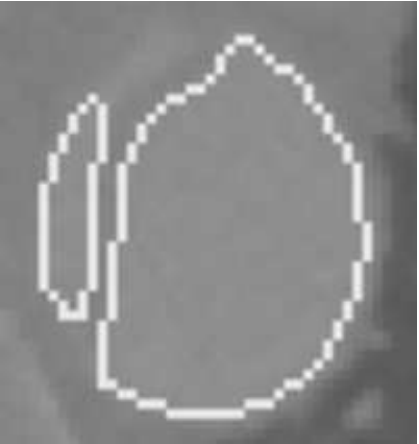}& 
\includegraphics[width=1.8cm]{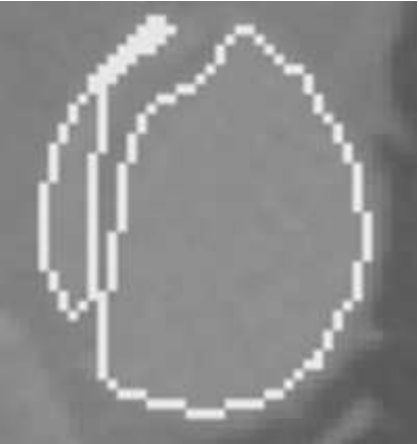}& 
\includegraphics[width=1.8cm]{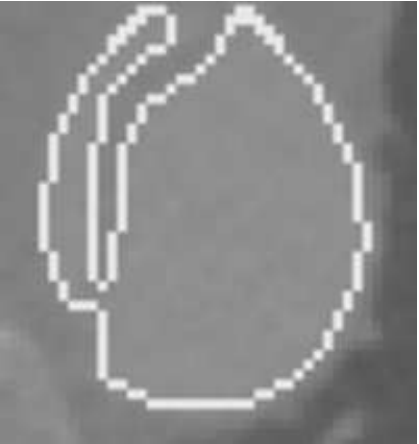}& 
\includegraphics[width=1.8cm]{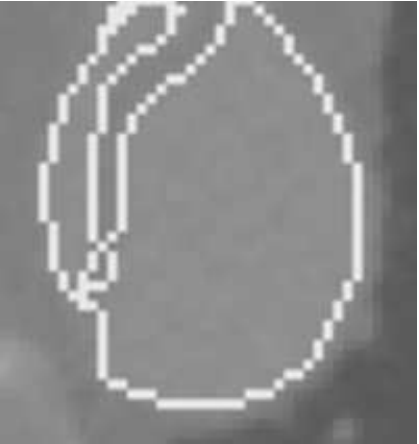}& 
\includegraphics[width=1.8cm]{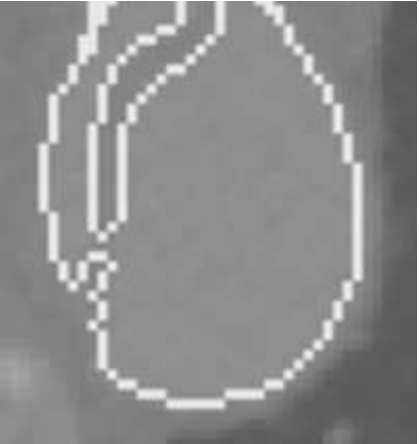}& 
\includegraphics[width=1.8cm]{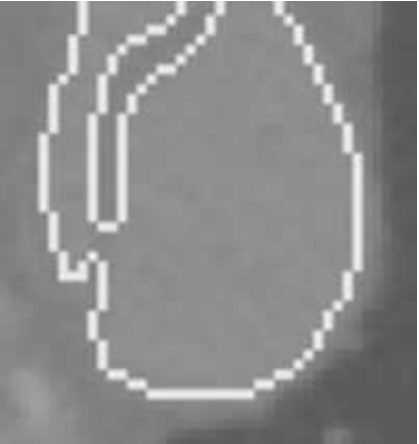}& 
\includegraphics[width=1.8cm]{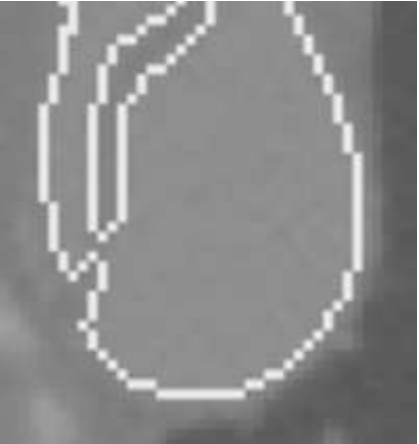}& 
\includegraphics[width=1.8cm]{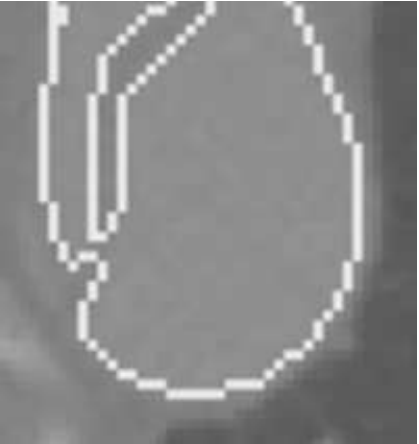}& 
\includegraphics[width=1.8cm]{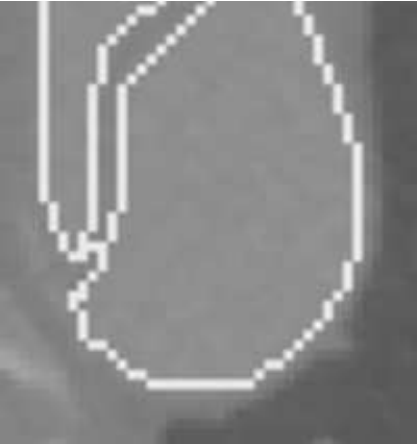}& 
\includegraphics[width=1.8cm]{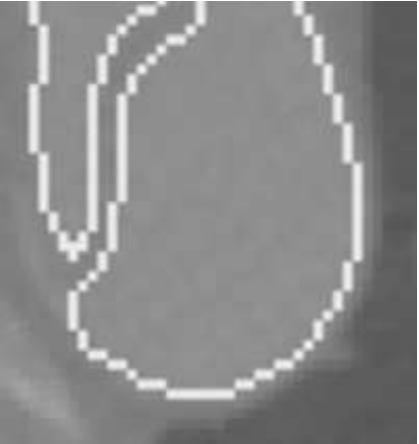}\\
\raisebox{0.9cm}{(f)} & \includegraphics[width=1.8cm]{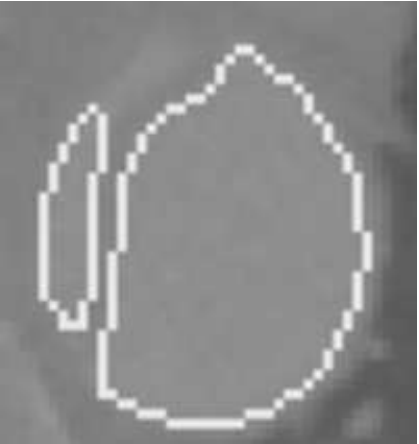}& 
\includegraphics[width=1.8cm]{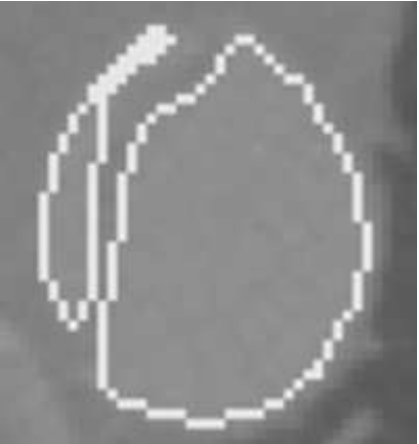}& 
\includegraphics[width=1.8cm]{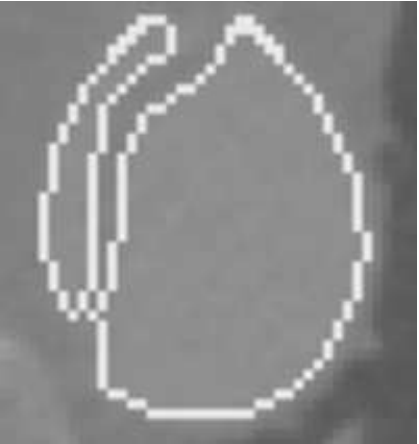}& 
\includegraphics[width=1.8cm]{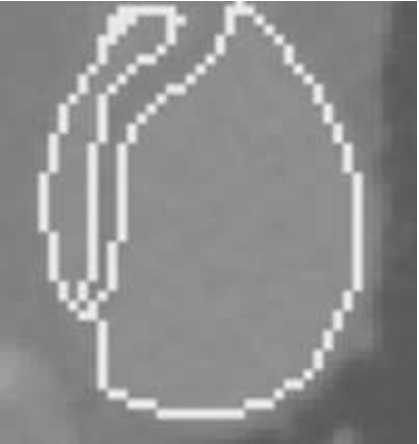}& 
\includegraphics[width=1.8cm]{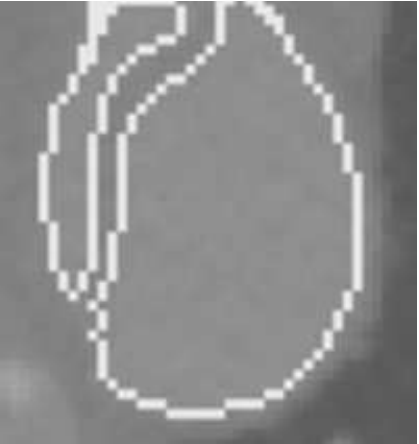}& 
\includegraphics[width=1.8cm]{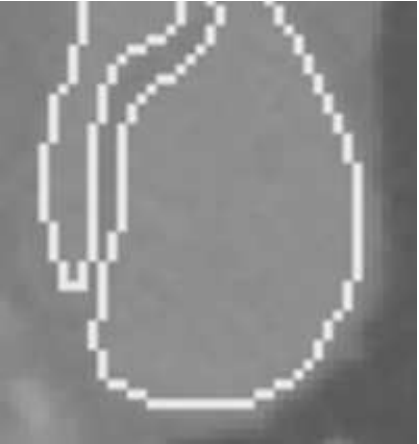}& 
\includegraphics[width=1.8cm]{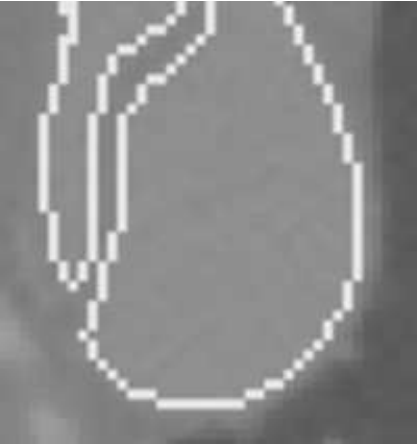}& 
\includegraphics[width=1.8cm]{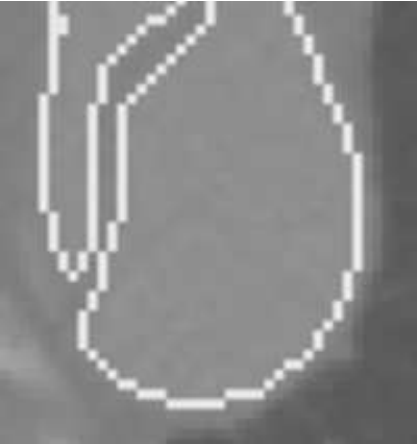}& 
\includegraphics[width=1.8cm]{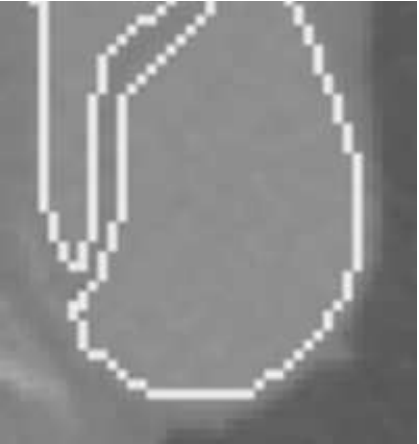}& 
\includegraphics[width=1.8cm]{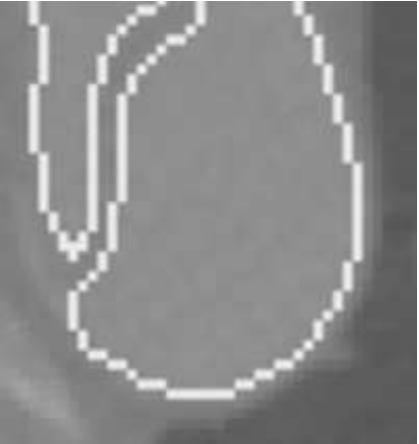}\\
\end{tabular}

\end{center}
\caption{Successive transverse sections (293-302). Contours (in white) of (a) connected lumens, (b) flap, (c) tear, (d) disconnected lumens, 
(e-f) zoom of (a) and (d) of the area shown by a square in (c)). We may observe in (f) the well disconnection of the 3D connected lumens of (e).}
\label{fig:fig11}
\end{figure}
\end{landscape}

\begin{figure}[t!]
\begin{center}
\begin{tabular}{ccccc}
\includegraphics[width=2.8cm, height=4cm]{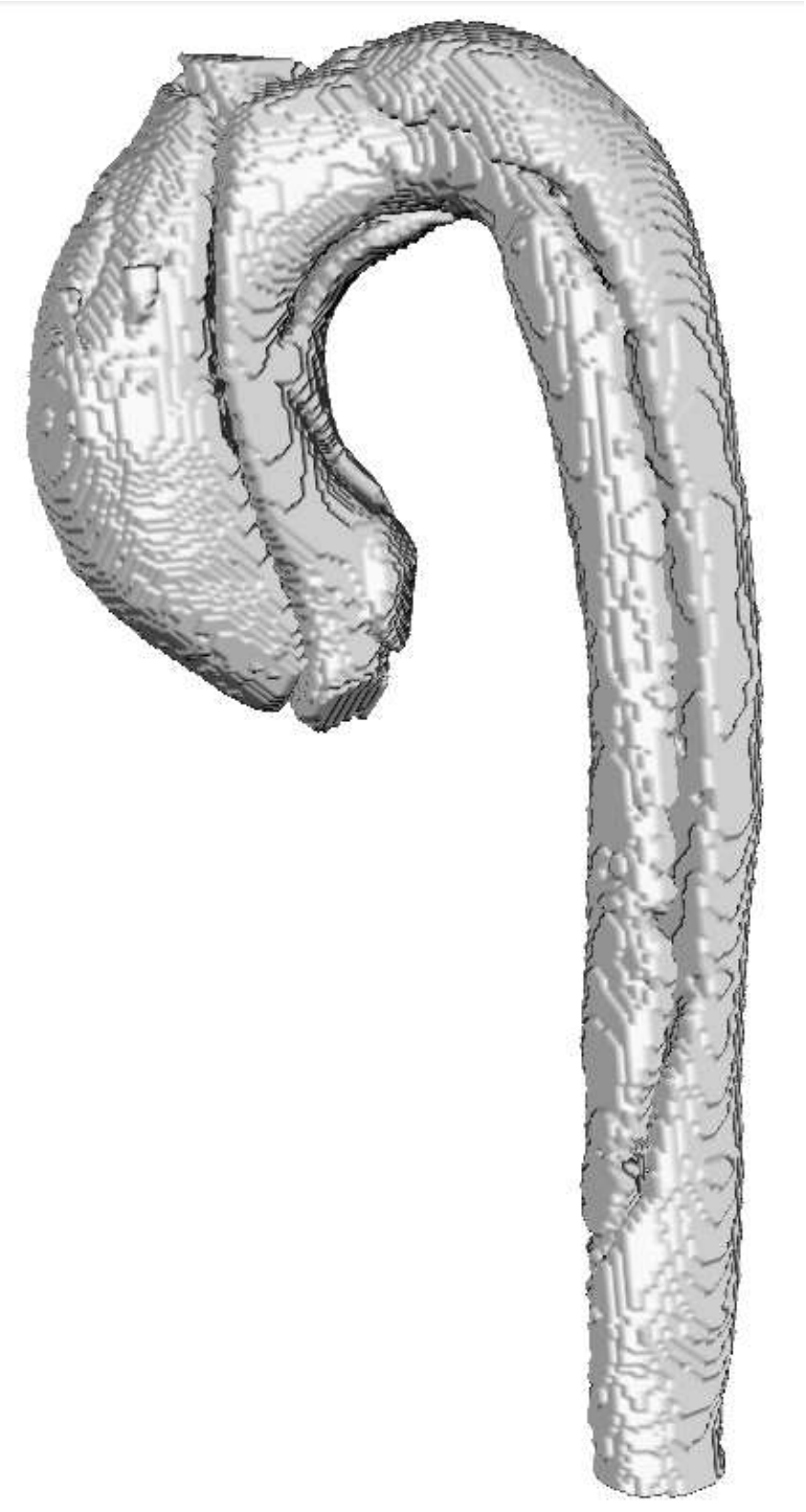} &
\includegraphics[width=2.8cm, height=4cm]{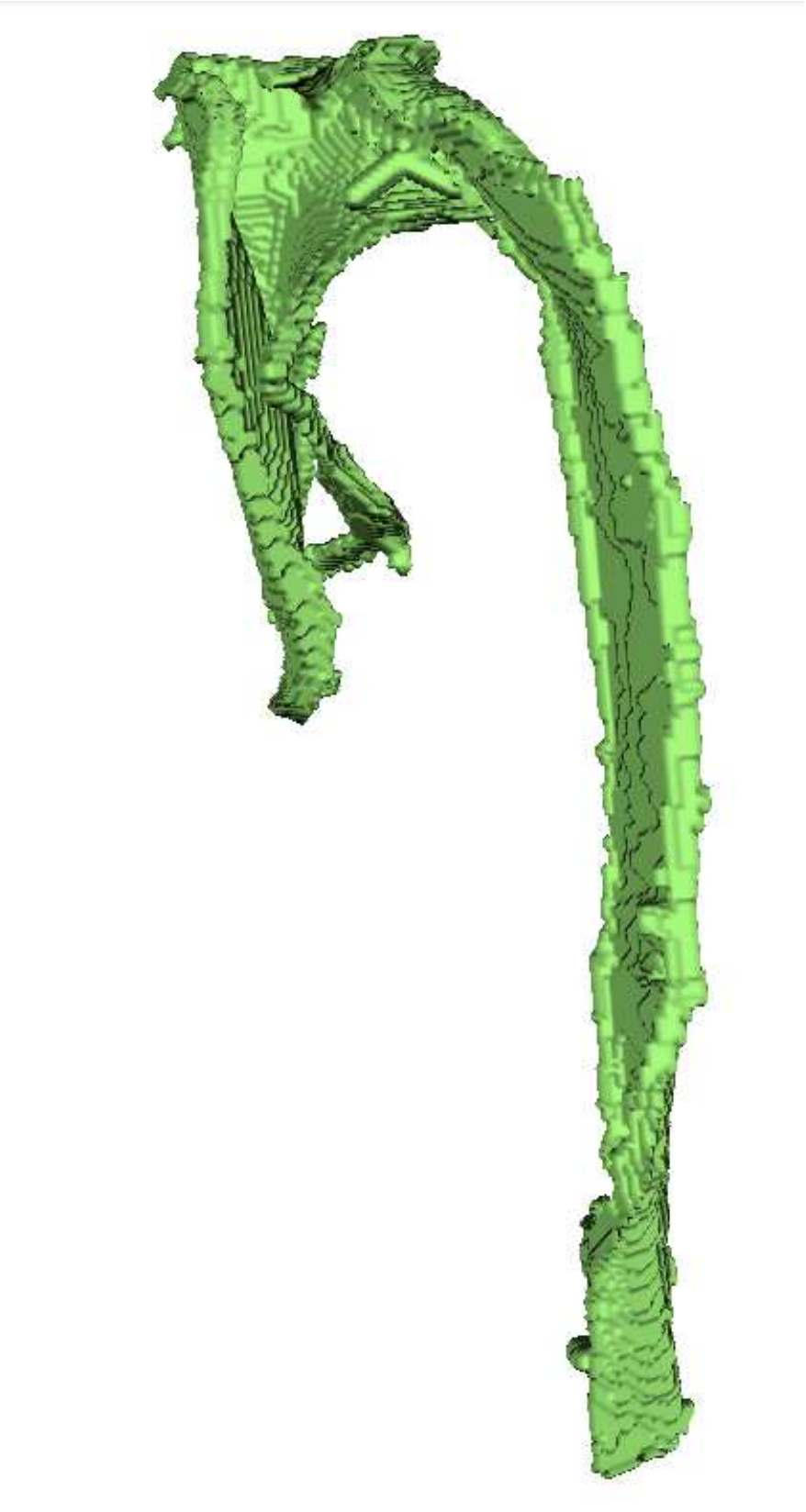} &
\includegraphics[width=2.8cm, height=4cm]{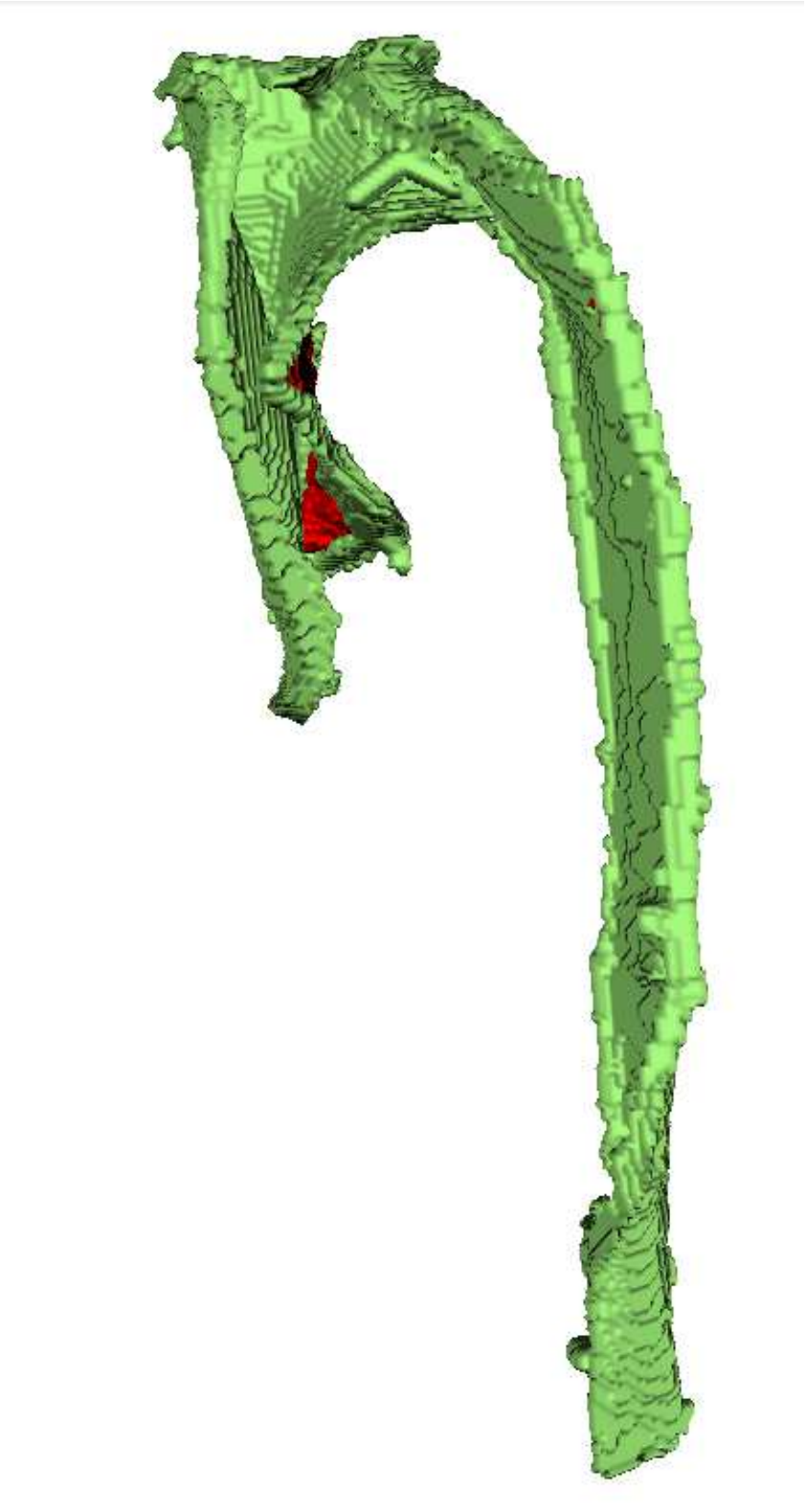} &
\includegraphics[width=2.8cm, height=4cm]{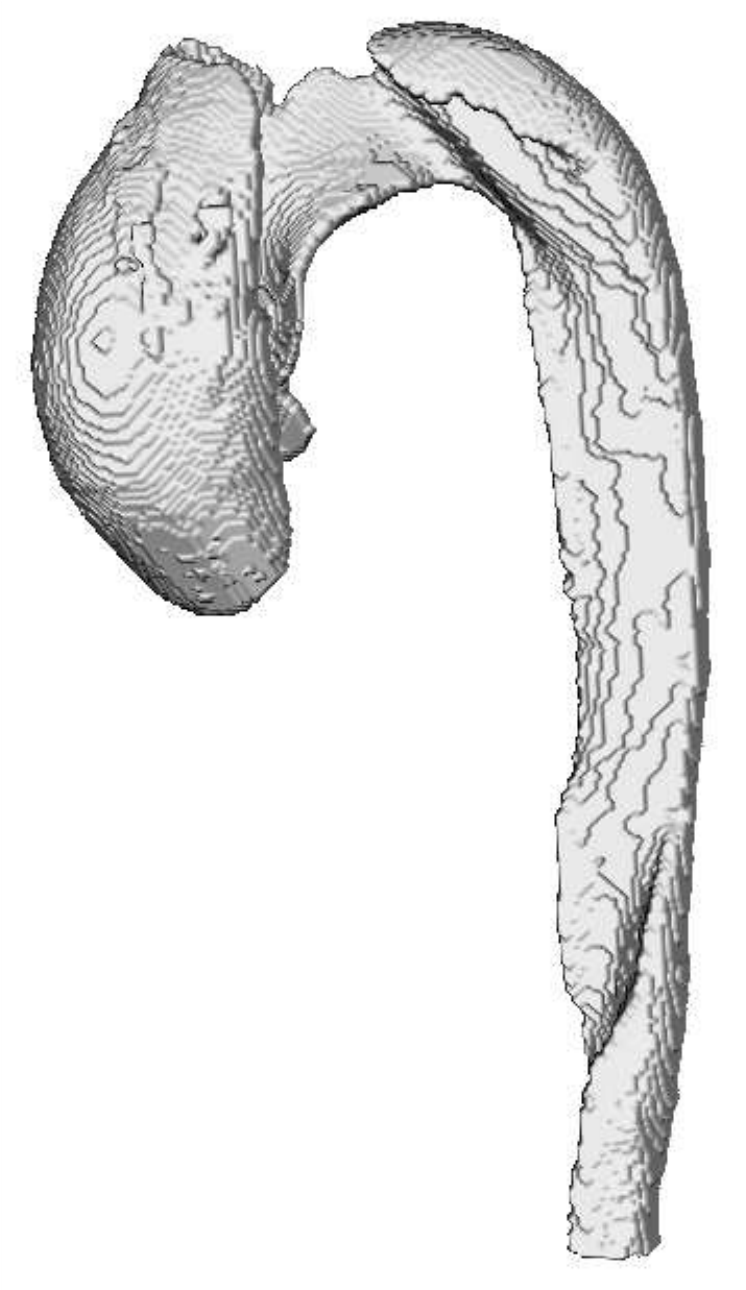} &
\includegraphics[width=2.8cm, height=4cm]{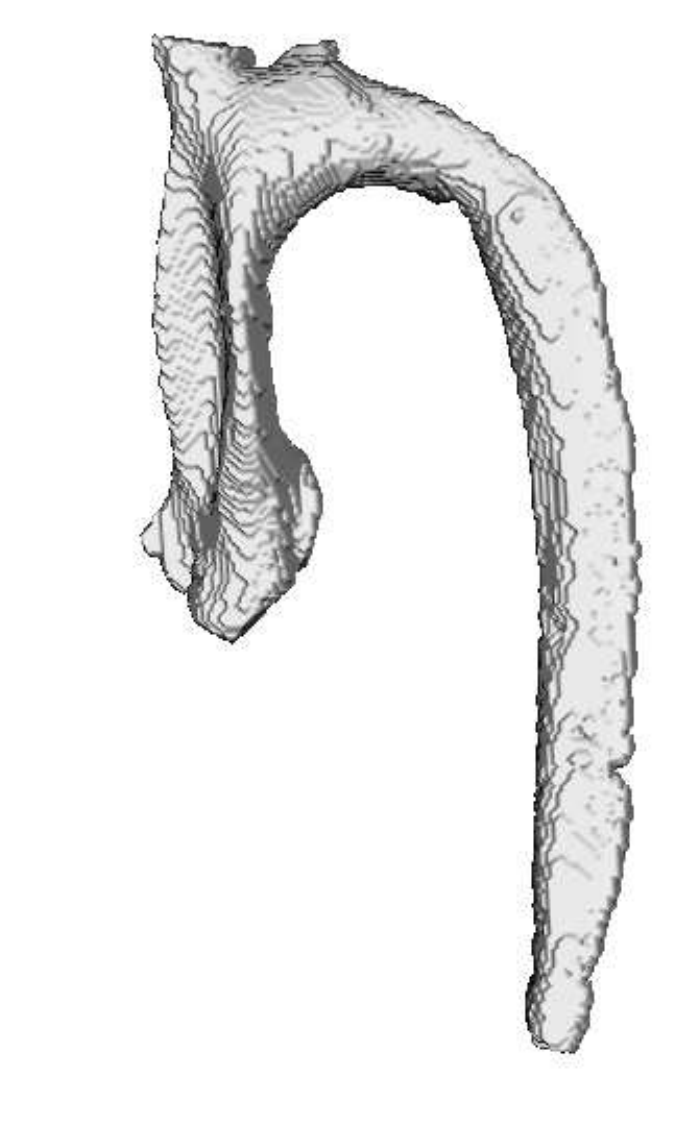} \\
(a) & (b) & (c) & (d) & (e) \\
\end{tabular}
\end{center}
\caption{(a) connected lumens, (b) flap, (c) addition of the main intimal tear in red, (d) first lumen, (e) second lumen.}
\label{fig:fig12}
\end{figure}

\begin{figure}[t!]
\begin{center}
\begin{tabular}{ccccc}
\includegraphics[width=2.8cm, height=4cm]{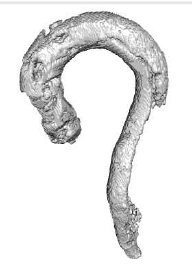} &
\includegraphics[width=2.8cm, height=4cm]{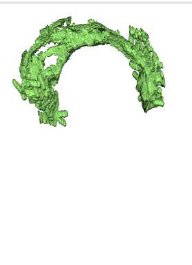} &
\includegraphics[width=2.8cm, height=4cm]{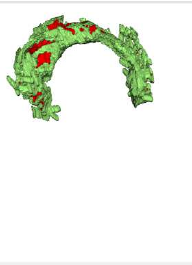} &
\includegraphics[width=2.8cm, height=4cm]{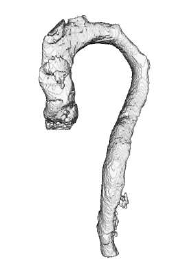} &
\includegraphics[width=2.8cm, height=4cm]{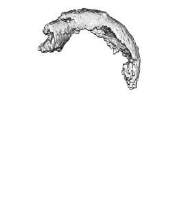} \\
(a) & (b) & (c) & (d) & (e) \\
\includegraphics[width=2.8cm, height=4cm]{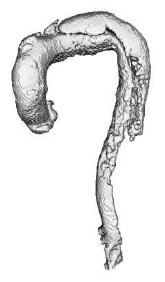} &
\includegraphics[width=2.8cm, height=4cm]{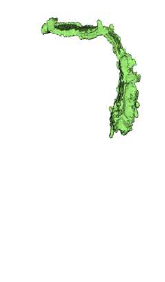} &
\includegraphics[width=2.8cm, height=4cm]{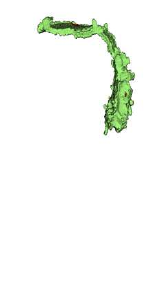} &
\includegraphics[width=2.8cm, height=4cm]{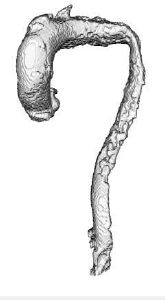} &
\includegraphics[width=2.8cm, height=4cm]{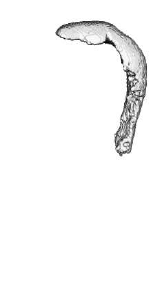} \\
(f) & (g) & (h) & (i) & (j) \\
\end{tabular}
\end{center}
\caption{(a, f) connected lumens, (b, g) flap, (c, h) addition of the main intimal tear in red, (d, i) first lumen, (e, j) second lumen, respectively for two other images of aortic dissection.}
\label{fig:fig13}
\end{figure}

Finally, in Fig. \ref{fig:fig12}, we propose the cartography of the current image (Fig. \ref{fig:fig9}-\ref{fig:fig11}) and in Fig. \ref{fig:fig13}, we give the results of the disconnection of the segmented connected lumens for two other images of aortic dissection 
by using the method proposed in this paper.

\section{Conclusion}

Using the Digital Topology, our previous works allowed us to locate the intimal tears and to place filling surfaces materializing such tears on the image of connected lumens. In this paper, we show how these surfaces allowed us to disconnect the image of segmented connected lumens. To our knowledge, this is the first time that surfaces filling holes are used to semi-automatically disconnect several parts of a 3D object. That also leads to propose a cartography of a dissection, it can be a valuable aid to physicians during their medical management of aortic dissections. As written in Sect. \ref{sec:context}, it would interesting to use these results, for example, to propose a registration of separated lumens in angiographic sequence \cite{Lohou2013b}, to study deformations of each lumen during cardiac and respiratory movements, to automatically distinguish between the true and false lumens, to use both connected and separated lumens in hemodynamics simulations.

\noindent \textbf{Author's contribution}\\
C.L.: conceptualization, methodology, formal analysis, investigation, writing, visualization, supervision.\\
B.M.: medical images.

\noindent \textbf{Acknowledgement}\\
The authors thank Guillaume Pascal (student at IUT, Le Puy-en-Velay, France) 
for the design of a graphical interface of medical images processing 
that yields several images of this paper.

\bibliographystyle{unsrt}
\bibliography{Article}

\begin{thebibliography}{10}

\bibitem{Lohou2013a}
C.~Lohou, N.~Fetnaci, P.~\L{}ubniewski, B.~Miguel, P.~Chabrot, and L.~Sarry.
\newblock Intimal flap segmentation on {CTA} aortic dissection images based on
  {M}athematical {M}orphology.
\newblock In {\em 6th Int. Congr. on Biomed. Eng. and Inform.}, pages 143--147,
  Hangzhou, China, 2013.

\bibitem{Aktouf2002}
Z.~Aktouf, G.~Bertrand, and L.~Perroton.
\newblock A three-dimensional holes closing algorithm.
\newblock {\em Pattern Recognit. Lett.}, 23(5):523--531, 2002.

\bibitem{Lohou2011}
C.~Lohou and B.~Miguel.
\newblock Detection of the aortic intimal tears by using 3d {D}igital
  {T}opology.
\newblock In {\em SPIE Electron. Imaging: Three-Dimensional Imaging, Interact.,
  and Meas.}, volume 7864, page 78640Z, San Francisco, USA, 2011.

\bibitem{Kovacs2006}
T.~Kov\'acs, P.~Cattin, H.~Alkadhi, S.~Wildermuth, and G.~Sz\'ekely.
\newblock Automatic segmentation of the aortic dissection membrane from 3d
  {CTA} images.
\newblock In G.-Z. Yang, D.~Shen, L.~Gu, and J.~Yang, editors, {\em Medical
  Imaging and Augmented Reality}, volume 4091 of {\em Lecture Notes in Computer
  Science}, pages 317--324. Springer Berlin Heidelberg, 2006.

\bibitem{Krissian2014}
K.~Krissian, J.~M. Carreira, J.~Esclarin, and M.~Maynar.
\newblock Semi-automatic segmentation and detection of aorta dissection wall in
  {MDCT} angiography.
\newblock {\em Med. Image Anal.}, 18(1):83--102, 2014.

\bibitem{Maldjian2012}
P.~D. Maldjian and L.~Partyka.
\newblock Intimal tears in thoracic aortic dissection: appearance on {MDCT}
  with virtual angioscopy.
\newblock {\em Am. J. of Roentgenol.}, 198(4):955--961, 2012.

\bibitem{Hossien2015}
A.~Hossien, S.~Gelsomino, B.~Mochtar, J.~G. Maessen, and P.~Sardari Nia.
\newblock Novel multi-dimensional modelling for surgical planning of acute
  aortic dissection type {A} based on computed tomography scan.
\newblock {\em Eur. J. of Cardiothorac. Surg.}, 48(5):e95--101, 2015.

\bibitem{Lee2008}
N.~Lee, H.~Tek, and A.F. Laine.
\newblock True-false lumen segmentation of aortic dissection using multiscale
  wavelet analysis and generative discriminative model matching.
\newblock In {\em SPIE Med. Imaging: Comput.-Aided Diagn.}, volume 6915, page
  69152V, San Diego, USA, 2008.

\bibitem{Mellissano2012}
G.~Melissano, L.~Bertoglio, E.~Rinaldi, E.~Civilini, Y.~Tshomba, A.~Kahlberg,
  E.~Agricola, and R.~Chiesa.
\newblock Volume changes in aortic true and false lumen after the {PETTICOAT}
  procedure for a type {B} aortic dissection.
\newblock {\em J. of Vasc. Surg.}, 55(3):641--651, 2012.

\bibitem{Suh2014}
G.-Y.~K. Suh, R.E. Beygui, D.~Fleischmann, and C.~P. Cheng.
\newblock Aortic arch vessel geometries and deformations in patients with
  thoracic aortic aneurysms and dissections.
\newblock {\em J. of Vasc. and Interv. Radiol.}, 25(12):1903--1911, 2014.

\bibitem{Li2018}
J.~Li, Y.~Ge, W.~Cheng, M.~Bowen, and G.~Wei.
\newblock Multi-task deep convolutional neural network for the segmentation of
  type {B} aortic dissection.
\newblock arXiv e-prints, arXiv:1806.09860, 2018.

\bibitem{Chen2013}
D.~Chen, M.~M\"{u}ller-Eschner, H.~von Tengg-Kobligk, D.~Barber,
  D.~B\"{o}ckler, R.~Hose, and Y.~Ventikos.
\newblock A patient-specific study of type-{B} aortic dissection: evaluation of
  true-false lumen blood exchange.
\newblock {\em Biomed. Eng. Online}, 12(65), 2013.

\bibitem{Menichini2016}
C.~Menichini, Z.~Cheng, R.G.J. Gibbs, and X.~Yun Xu.
\newblock Predicting false lumen thrombosis in patient-specific models of
  aortic dissection.
\newblock {\em J. of the Royal Soc. Interface}, 13(124):20160759, 2016.

\bibitem{DillonMurphy2016}
D.~Dillon-Murphy, A.~Noorani, D.~Nordsletten, and C.~Alberto Figueroa.
\newblock Multi-modality image-based computational analysis of haemodynamics in
  aortic dissection.
\newblock {\em Biomech. and Model. in Mechanobiol.}, 15(4):857--876, 2016.

\bibitem{Tatco}
V.~Tatco.
\newblock Pathogenesis of aortic dissection.
\newblock
  https://radiopaedia.org/cases/pathogenesis-of-aortic-dissection-illustration,
  2016.
\newblock Case courtesy of Dr Vincent Tatco, Radiopaedia.org, rID: 48452,
  Creative Commons CC-BY-NC-ND.

\bibitem{Adams94}
R.~Adams and L.~Bischof.
\newblock Seeded region growing.
\newblock {\em IEEE Trans. on Pattern Anal. and Mach. Intell.}, 16(6):641--647,
  1994.

\bibitem{McInerney96}
T.~McInerney and D.~Terzopoulos.
\newblock Deformable models in medical image analysis: a survey.
\newblock {\em Med. Image Anal.}, 1(2):91--108, 1996.

\bibitem{Serra88}
J.~Serra.
\newblock {\em Image Analysis and Mathematical Morphology. Volume II:
  theoretical advances}.
\newblock Acad. Press, 1988.

\bibitem{Lohou2005}
C.~Lohou and G.~Bertrand.
\newblock A 3d 6-subiteration curve thinning algorithm based on {$P$}-simple
  points.
\newblock {\em Discret. Appl. Math.}, 151(1-3):198--228, 2005.

\bibitem{Kong1989}
T.Y. Kong and A.~Rosenfeld.
\newblock {D}igital {T}opology: introduction and survey.
\newblock {\em Comput. Vis., Graph., and Image Process.}, 48(3):357--389, 1989.

\bibitem{Dijkstra1959}
E.W. Dijkstra.
\newblock A note on two problems in connection with graphs.
\newblock {\em Numerische Mathematik}, 1(1):269--271, 1959.

\bibitem{Fetnaci2013}
N.~Fetnaci, P.~\L{}ubniewski, B.~Miguel, and C.~Lohou.
\newblock 3d segmentation of the true and false lumens on {CT} aortic
  dissection images.
\newblock In {\em SPIE Electron. Imaging: Three-Dimensional Image Process.
  (3DIP) and Appl.}, volume 8650, page 86500M, Burlingame, USA, 2013.

\bibitem{Bertrand1994}
G.~Bertrand.
\newblock Simple points, topological numbers and geodesic neighborhoods in
  cubic grids.
\newblock {\em Pattern Recognit. Lett.}, 15(10):1003--1011, 1994.

\bibitem{Lubniewski2012}
P.~\L{}ubniewski, B.~Miguel, V.~Sauvage, and C.~Lohou.
\newblock Interactive 3d segmentation by tubular envelope model for the aorta
  treatment.
\newblock In {\em SPIE Electron. Imaging: Three-Dimensional Image Process.
  (3DIP) and Appl. II}, volume 8290, page 82901F, Burlingame, USA, 2012.

\bibitem{Wolf2005}
I.~Wolf, M.~Vetter, I.~Wegner, T.~B\"{o}ttger, M.~Nolden, M.~Sch\"{o}binger,
  M.~Hastenteufel, T.~Kunert, and H.-P. Meinzer.
\newblock The medical imaging toolkit.
\newblock {\em Med. Image Anal.}, 9(6):594--604, 2005.

\bibitem{Lohou2013b}
C.~Lohou, P.~\L{}ubniewski, N.~Fetnaci, H.~Feuill{\^a}tre, J.~Courbon,
  V.~Sauvage, J.-Y. Boire, L.~Boyer, L.~Camilleri, L.~Cassagnes, P.~Chabrot,
  B.~Miguel, and L.~Sarry.
\newblock Interventional planning and assistance for ascending aorta
  dissections.
\newblock {\em IRBM}, 34(4-5):306--310, 2013.

\end{thebibliography}

\end{document}